\documentclass[%
 reprint,
nofootinbib,
 amsmath,amssymb,
 aps,
]{revtex4-2}
\usepackage{subfig}
\usepackage{tabularx}
\usepackage{threeparttable}
\usepackage{booktabs}
\usepackage{graphicx}
\usepackage{dcolumn}
\usepackage{bm}
\usepackage{hyperref}

\RequirePackage[bottom]{footmisc}
\usepackage{xcolor}

\DeclareUnicodeCharacter{2212}{\textendash}

\RequirePackage{tikz}

\newcommand{\aap}{Astron. Astrophys.}
\newcommand{\apjl}{Astrophys. J. Lett.}
\newcommand{\apjs}{Astrophys. J. Suppl. Ser.}
\newcommand{\aj}{Astron. J.}
\newcommand{\mnras}{Mon. Not. R. Astron. Soc.}
\newcommand{\pasj}{Publ. Astron. Soc. Jpn.}
\newcommand{\pasp}{Publ. Astron. Soc. Pac.}
\newcommand{\araa}{Annu. Rev. Astron. Astrophys.}
\newcommand{\ssr}{Space Sci. Rev.} 

\newcommand{\jcap}{J. Cosmol. Astropart. Phys.}
\newcommand{\physrep}{Phys. Rep.}

\begin{document}

\preprint{APS/123-QED}

\title{Excess of diffuse gamma-ray emission detected from the galaxy cluster Abell 119 from 14-year \textit{\textit{Fermi}}-LAT Data \protect\thanks{Gamma ray excess in Abell 119}}

\author{Gajanan D Harale}\email{gajanan.harale@fergusson.edu}
\affiliation{Department of Physics, Savitribai Phule Pune University, Pune - 411007, Maharashtra, India}
 \altaffiliation[Also at ]{Department of Physics, Fergusson College (Autonomous), Pune - 411004, Maharashtra, India}
\author{Surajit Paul}%
 \email{surajit.paul@manipal.edu}
\affiliation{Manipal centre for Natural Sciences, Manipal Academy of higher education, Manipal - 576104, Karnataka, India}%

\date{\today}

\begin{abstract}
Galaxy clusters are the vast and massive gravitationally bound structures in the Universe and are often considered as the largest reservoirs of high-energy cosmic ray particles. However, no direct and conclusive detection of $\gamma$-rays from them has been reported to date. This would mean either our understanding of these systems is insufficient or it could be due to the limited sensitivity and poor localization capabilities of $\gamma$-ray telescopes. Moreover, strong interactions of $\gamma$-rays with intergalactic material along the line of sight allow detectable signals from only a few very nearby clusters. The detection prospects are further constrained to dynamically active clusters, which are theoretically expected to produce detectable $\gamma$-ray emissions. Given these constraints, 
we systematically shortlisted a suitable sample of nearby ($z < 0.05$), dynamically active galaxy clusters and performed a comprehensive analysis of 14 years of Fermi-LAT data. In this work, we present a detailed study of diffuse $\gamma$-ray emission around one such galaxy cluster, Abell 119 (A119). Known for its ongoing merger activity, A119 exhibits significant X-ray luminosity and complex dynamical activity. We analyzed the data using the \textit{Fermipy} package and \textit{Fermi} Science Tools for spatial and spectral models to account for all potential sources of $\gamma$-ray emission. Our analysis confirms all the 4FGL point sources $4FGL J0059.3-0152$, $4FGL J0101.0-0059$, and $4FGL J0059.2+0006$ with significant TS values. It further reveals a $\sim 4\sigma$ excess of diffuse $\gamma$-ray emission, offset by $\sim0.25^{\circ}$ from the cluster center - suggesting a possible association with the cluster halo region. The extended model provides the most robust description for the detected $\gamma$-ray excess. The computed luminosity bounds for the extended models $\sim 12.21^{+2.74}_{-3.95}$$\times10^{42}$ erg s$^{-1}$ and the particle spectrum $\sim 2.25^{+0.38}_{-0.13}$ are consistent with previous studies and provide further evidence for the presence of diffuse gamma-rays in this galaxy cluster. 
With the available information, we conclude that the observed emission may arise from non-thermal processes in the intracluster medium (ICM), most likely 
hadronic process. Although the observed $\sim 4\sigma$ signal is significant, uncertainties in source localization and other parameters and instrumental limitations prevent us from confirming the detection. Nonetheless, our results underscore the potential for deeper exploration of $\gamma$-rays in the cluster environment with improved sensitivity and resolution. Moreover, an estimated neutrino flux $E^{2}\phi_{\nu} \approx 3 \times 10^{-10} \ \mathrm{GeV  ~cm^{-2} s^{-1} sr^{-1}}$  telescope provides motivation for future observations with upcoming neutrino telescopes.

\end{abstract}

\maketitle


\section{Introduction}\label{sec1}

Galaxy clusters are the largest self-gravitating structures known in the Universe. They comprise tens to hundreds of galaxies and the intergalactic medium, all gravitationally bound within megaparsec-scale dark matter halos \cite{Bykov2015SSRv, Yu2015ApJ, Springel2006Natur}. The energy budget of these clusters is regulated by various dynamical processes, including galaxy formation, galactic winds, active galactic nucleus (AGN) activity, star formation, supernova explosions, and merger activities across different scales (galaxy, group, and cluster levels). Although the intracluster medium is predominantly composed of hot gas emitting thermal radiation, a non-negligible fraction of the cluster energy is non-thermal in nature,  comprising cosmic rays, gamma rays, and other high-energy particles \citep{Brunetti_2014IJMPD, Petrosian2008SSRv}. Particle acceleration induced by dynamical events, as well as dark matter (DM) annihilation, has been proposed as a primary mechanism for the generation of these high-energy particles in clusters \citep{Planck_Collaboration2013, Huber2013A&A, Berezinsky1997ApJ}.

The galaxy clusters are generally dynamically active as they form through continuous accretion and mergers of small to large groups of galaxies \cite{West1995}. Theoretical models, as well as, cosmological simulations show that the cluster merger events are associated with a large amount of energy release ($\sim10^{64}~\rm{erg~s^{-1}}$ \citep{Sarazin_2002ASS}). The released energy then dissipates in the cluster medium, primarily through the formation and propagation of Mpc scale shocks and injection of turbulent motions in the intracluster medium \citep{Rottgering1997, Paul_2011ApJ}. These strong collision-less processes are known to produce high-energy relativistic cosmic-ray particles (CRs) via diffusive shock acceleration (DSA; \citealt{Bladford1987}) or stochastically through turbulent re-acceleration or \textit{Fermi}-II process \citep{Petrosian2008SSRv}. This makes merging clusters of galaxies one of the best sites for studying high-energy cosmic ray particles, especially $\gamma$-rays that provide direct evidence of the presence of cosmic rays in the cluster medium.

Astrophysical acceleration processes, as mentioned above, can transfer a significant fraction of energy to cosmic ray (CR) protons. These highly energetic charged particles are expected to interact with the thermal protons of the intracluster medium (ICM), leading to the production of one neutral pion along with two charged pions. The subsequent decay of neutral pions generates high-energy $\gamma$-rays \cite{Stecker1973ApJ, Aharonian2008, Kelner2006}, contributing substantially to the cluster’s $\gamma$-ray emission, and charged pion decay into corresponding neutrino particles. This mechanism is known as the hadronic process and is considered a primary channel for high-energy $\gamma$-ray production in clusters \cite{Aharonian2009A&A, HESSCollaboration2012A&A, Hussain2023}. These proton acceleration processes are also expected to accelerate electrons, creating a population of high-energy CR electrons in the cluster. CR electrons may also originate from active galactic nuclei (AGN) activity, stellar winds, supernova explosions, and other sources within the cluster environment. These energetic CR electrons can further up-scatter low-energy photons - primarily from the cosmic microwave background (CMB) - via inverse-Compton (IC) scattering, boosting them to $\gamma$-ray energies \cite{Blasi1999}. This leptonic process thus represents another potential contributor to the high-energy $\gamma$-ray emission from the ICM.

While, the presence of CR electrons has been confirmed through the detection of hard X-rays \citep{Petrosian2008SSRv} and diffuse radio synchrotron emissions \citep[e.g.,][]{Buote2001, Donnert2011PhDT, Feretti2012, Osinga2021A&A}, a corresponding diffuse $\gamma$-ray signal from galaxy clusters remains elusive \citep{Wittor2016, Vazza2023A&A}. In this context, a longstanding debate persists in the literature concerning the origin of diffuse cluster emissions -- particularly radio halos and mini-halos. Discussions continue regarding their hadronic versus leptonic origins \citep{Brunetti_2014IJMPD, Weeren2019SSRv, Donnert2010}. A conclusive detection of diffuse $\gamma$-rays and the corresponding neutrinos from clusters would serve as an important key observational constraint to distinguish between these models.

As we have noticed, both the hadronic and the leptonic processes require the presence of high-energy CR protons or electrons. To produce the required pull of CR particles, merging activities in the cluster is essential. To have enough thermal protons for the hadronic process, the cluster should be hotter than the usual \citep{Guo2008MNRAS}. Furthermore, because of high attenuation of $\gamma$-rays and to have enough resolution of the images to resolve and distinguish the diffuse emission from the point sources, low redshift objects would be preferred. With this in mind, we have shortlisted a number of low-redshift and dynamically active clusters and performed a thorough analysis of 14 years of \textit{Fermi} data for a few of these sources. 

In this paper, we present our analysis of one of the shortlisted clusters, Abell 119 (hereafter A119). The cluster exhibits signatures of ongoing mergers in multi-wave band observations, indicating possibility of detection of diffuse gamma-ray emissions in \textit{Fermi}-LAT data. By focusing on A119, we aim to deepen our understanding of the non-thermal processes and interactions between cosmic rays and intracluster medium, which may play a critical role in shaping the high-energy landscape of such massive structures in the Universe.

We used a standard $\Lambda$CDM cosmology with $H_0 = 67.74$ km s$^{-1}$ Mpc$^{-1}$, $\Omega_M = 0.3$, and $\Omega_{ \Lambda} = 0.7$. At z $= 0.044$ \citep{Smith2004},$R_{500}=1140$ kpc, the luminosity distance is $D_L= 201.3$ Mpc, the angular size distance is $D_A =184.7$ Mpc throughout this paper.

\section{Cluster Abell 119}

A119 is one of the nearest and massive merging clusters (redshift z$=0.044$  \citep{Smith2004}) with a mass of $3.05\times 10^{14}M_{\odot}$ \citep{Way1997}. The cluster is particularly interesting because of its high X-ray luminosity \citep{Lee2016}, approximately $3\times 10^{44}$erg s$^{-1}$\citep{Edge1990,David1993,Markevitch1998}.
It has a history of off-axis mergers \citep{Watson2023} and displays a dynamically complex structure, with obvious lack of an X-ray cool core \citep{Lagan2019}. The presence of substructure within its field confirms its ongoing merger activities \citep{Sreeoh2016,Watson2023}.

\section{\textit{Fermi}-Lat Data Analysis}\label{sec2}

\begin{figure*}
    \subfloat[100 MeV-1 TeV]{%
    \includegraphics[width=0.32\textwidth]{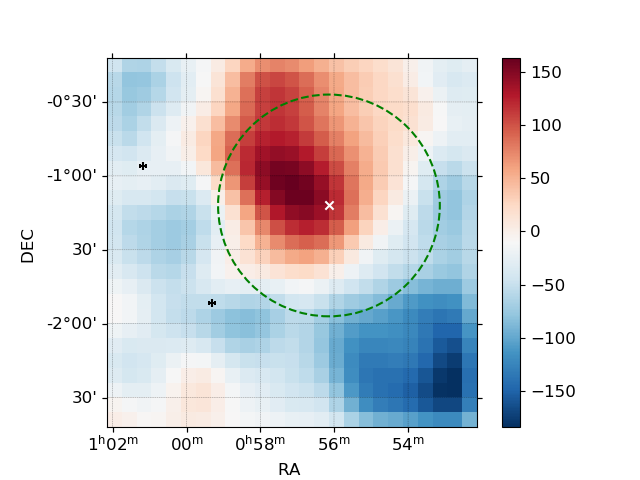}%
        }%
    \hfill%
    \subfloat[100 MeV-1 GeV]{%
        \includegraphics[width=0.32\textwidth]{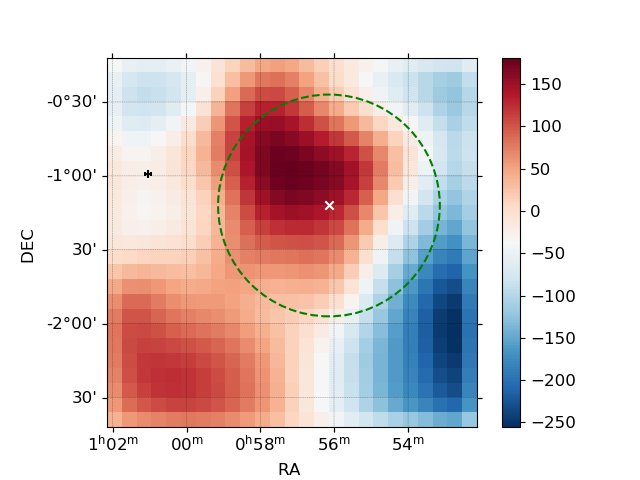}%
        }%
    \hfill%
    \subfloat[1 GeV-1 TeV]{%
        \includegraphics[width=0.32\textwidth]{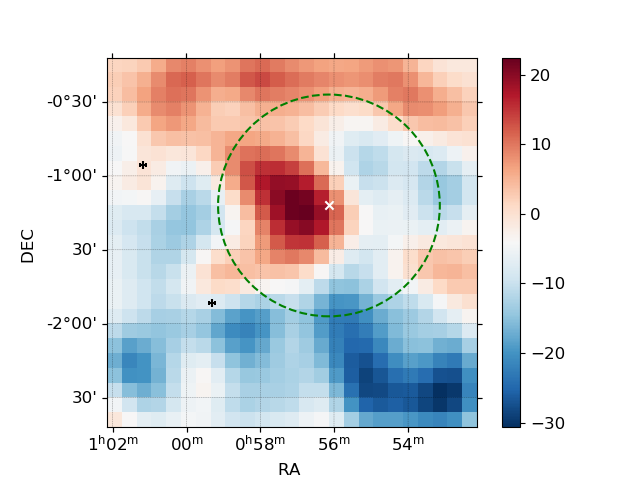}%
        }%
    \caption{ Residual excess map in the vicinity of the A119 cluster for the total energy band $100$ MeV - $1$ TeV, $100$ MeV - $1$ GeV, and $1$ GeV - $1$ TeV. The dotted green circle shows the extension of the $\gamma$-ray emission, the white cross indicates the center of the cluster, and the black $+$ symbols represent 4FGL-DR4 sources within the region.}
    \label{fig:residual map}  
\end{figure*}

We have selected the \textit{Fermi}-LAT Pass 8 data from August 4, 2008 (MET 239557417)  to March 14, 2024 (MET 732104868) for A119 with RA=$14.0357^{\circ}$, Dec.=$-1.20021^{\circ}$ \citep{Piffaretti2011} and with radius $20^{\circ}$ in the photon energy range $100$ MeV to $1$ TeV and obtained it from \textit{Fermi}-LAT server\footnote{\url{https://Fermi.gsfc.nasa.gov/ssc/data}}. We analyzed the data using \textit{Fermipy}\footnote{\url{https://github.com/FermiPy/Fermipy}} version v1.2~\citep{Wood2017} and \textit{\textit{Fermi} Science Tools version} 2.2.0\footnote{\url{https://Fermi.gsfc.nasa.gov/ssc/data/analysis/software/}}. For further analysis, we have selected and binned the data for the region of interest (ROI), a square region of $12^{\circ}\times 12^{\circ}$ with the center at RA=$14.0357^{\circ}$, DEC=$-1.20021^{\circ}$ \citep{Piffaretti2011} as the region of interest (ROI). P8R3$\_$SOURCE$\_$V3, the instrument response functions (IRFs) used to analyze the events in the ROI with evtype=3 and evclass =128. We also used the recommended data quality criteria (DATA$\_$QUAL $>$ 0)\&\&(LAT$\_$CONFIG == 1) to perform the (GTIs) good time intervals which are in the spacecraft file. To account for the $\gamma$-ray contamination from the Earth's limb, we used a zenith angle of $90^{\circ}$. For binned analysis, we adopted a maximum-likelihood optimization technique. The recently released \textit{Fermi}-LAT 14-year catalog 4FGL-DR4  \cite{Ballet2023} was used, which includes point-like sources and extended sources in the background model within the region of a radius of $12^{\circ}$ with latest Galactic diffuse emission model \textit{gll$\_$iem$\_$v07.fits} to account for the diffuse background component and isotropic extragalactic emission model \textit{iso$\_$P8R3$\_$SOURCE$\_$V3$\_$v1.txt}.

This study utilizes a sophisticated likelihood analysis technique, which leverages advanced statistical methodologies to rigorously examine the presence, distribution, and characterize the $\gamma$-ray emissions from the galaxy cluster A119. To investigate the residual excess $\gamma$-ray signal linked to this object, as illustrated in \autoref{fig:residual map}, we applied three types of models, point source model,  morphological models and CRp models (radial models which consists of spatial and spectral properties).

\section{Overview of the used models}
 
\subsection{The point source model}
Massive galaxies in the cluster are expected to reside near the cluster centre. The excess of $\gamma$-ray emission in the vicinity of cluster A119 was thus first modelled using the point source model. We incorporated the point source model at the cluster centre RA=$14.0357^{\circ}$, Dec.=$-1.20021^{\circ}$ into our background model. Subsequently, we performed fitting using the \textit{fit} function of \textit{Fermipy}. Following the fitting, we utilized the \textit{localize} function of \textit{Fermipy} to determine the best-fit position, which was found to be at RA, DEC =$14.3522^{\circ}\pm0.3647^\circ$,$-1.6548^{\circ}\pm0.3217^\circ$ with  offset $= 0.5538^\circ$, r68 $=   0.5104^\circ$, r95 $= 0.8237^\circ$, r99 $=1.0214^\circ$ for A119.

\subsection{Spatial Morphological Models} Mophological models analyze the spatial distribution of the $\gamma$-ray signal. By examining the morphology, we can assess whether the emission originated from specific substructures within the cluster or from the diffuse intracluster medium. This helps in identifying potential sources and understanding how $\gamma$-ray emissions spatially correlated to known features of a cluster. Here, to analyze the cluster A119, we used spherical models such as radial Gaussian and disc model.

Furthermore, to refine our analysis, we incorporated complementary multi-wavelength data of A119. This multiwavelength approach aids both in comparing the $\gamma$-ray signal across other energy bands and in constructing templates to model the emission mechanism. We used the Planck telescope thermal Sunyaev-Zel'dovich data, X-ray data from ROSAT and XMM Newton, Sloan Digital Sky Survey (SDSS) optical data. Further details are provided in the following subsections.

\subsubsection{Radial Gaussian and Disc Model}
After the removal of all possible point sources, the origin of the residual excess diffuse emission of $\gamma$-ray signal was investigated within the $R_{500}$ of the cluster. Considering the usual spherical morphology of the cluster, we used the spherical models such as radial Gaussian model and Disc model. These models Were added at the cluster centre RA=$14.0357^{\circ}$ , Dec.=$-1.20021^{\circ}$, by varying the radii from $0.1^{\circ}$ to $0.7^{\circ}$ with a step size of $0.05^{\circ}$. Furthermore, the \textit{fit} function of \textit{Fermipy} was used to compute the TS value for each radius. We found that both models are very well fitted with radius $0.5^{\circ}$. By fitting the radial Gaussian and disc model, we aimed to identify the radial distribution and the extension of the $\gamma$-ray emission from the cluster halo, providing insights into the potential sources and mechanisms contributing to the observed emission.

\subsubsection{Galaxy Density Map}
While working on the morphological models with optical galaxy data, we first constructed a galaxy density map as shown in upper left panel of \autoref{fig:morphology check} for the A119 region using spectroscopic information from the SDSS database \footnote{\url{https://skyserver.sdss.org/CasJobs}}
dr18 data \footnote{\url{https://skyserver.sdss.org/dr18}}. To achieve this, galaxies were selected based on their redshift, and each galaxy is weighted according to its redshift distance from the cluster redshift $z_{A119}$, using a Gaussian weighting function $w=exp\left(\frac{-(z_{gal}C-z_{A119}C)^2}{2\sigma_V^2}\right)$, where $\sigma_V=778$ Km/s \cite{Feretti_1999}, C is the speed of light, $z_{gal}$ is the redshift of the galaxy and $z_{A119}$ is the redshift of the cluster of interest. The galaxies were then binned into a spatial grid, and the resulting density map is smoothed using a Gaussian kernel with a FWHM of 10 arcminutes to improve the signal-to-noise ratio. After smoothing, the background was estimated by averaging the map at radii greater than 60 arcminutes, beyond the cluster size, from the cluster center and subtracted to isolate the signal from the cluster itself. This method allows us to minimize the influence of background galaxies and focus on those dynamically associated with the cluster A119.

\subsubsection{Planck tSZ Data}
The thermal Sunyaev-Zel'dovich (tSZ) effect, described initially by Sunyaav \& Zel'dovich, (1972) \cite{SZ_effect}, is a powerful probe for analyzing clusters of galaxies, as it provides a measure of the integrated thermal pressure of the intracluster gas along the line of sight. This is crucial for comparing to the $\gamma$-ray signal. We utilized the NILC (Needlet Internal Linear Combination) Compton parameter map constructed from the Planck PR4 (NPIPE) data \citep{Planck_Collaboration_2020} described in \cite{NILC_2024} to achieve a adequate signal-to-noise image of the tSZ map from the cluster A119 with a resolution of 10 arcminutes. The Planck Legacy Archive (PLA) offers publicly available data \footnote{\url{http://pla.esac.esa.int/pla/}}.

\subsubsection{X-ray Data}

The diffuse X-ray emission due to thermal bremsstrahlung is the direct tracer of the distribution of intracluster gas \citep{x_ray_thermal_gas_density}. Since intracluster medium supplies the energetic protons and electrons that are possible sources of $\gamma$-ray emission \citep{Aharonian2008}, morphological correlation with X-rays can provide better understanding of the origin of the excess $\gamma$ emission from clusters. To prepare the data \citep[see for more details][]{MINOT2020}\footnote{\url{https://github.com/remi-adam/minot}}, we performed background subtraction and normalized the images using ROSAT exposure maps collected from ROSAT All Sky Survey (RASS)\footnote{\url{http://cade.irap.omp.eu/dokuwiki/doku.php?id=rass}}, ensuring accurate measurement of the diffuse X-ray emission associated with the cluster, we also incorporated the Rosat PSPC data \footnote{\url{https://heasarc.gsfc.nasa.gov/FTP/rosat/data/pspc/processed_data/}} \citep{Sanders_2000}, \citep{mass_temp_rosat2001}. The final image focuses on the diffuse cluster emission, facilitating a clearer comparison with $\gamma$-ray signal and other observational data.

\begin{figure*}[ht] 
    \subfloat[]{%
        \includegraphics[width=0.5\textwidth]{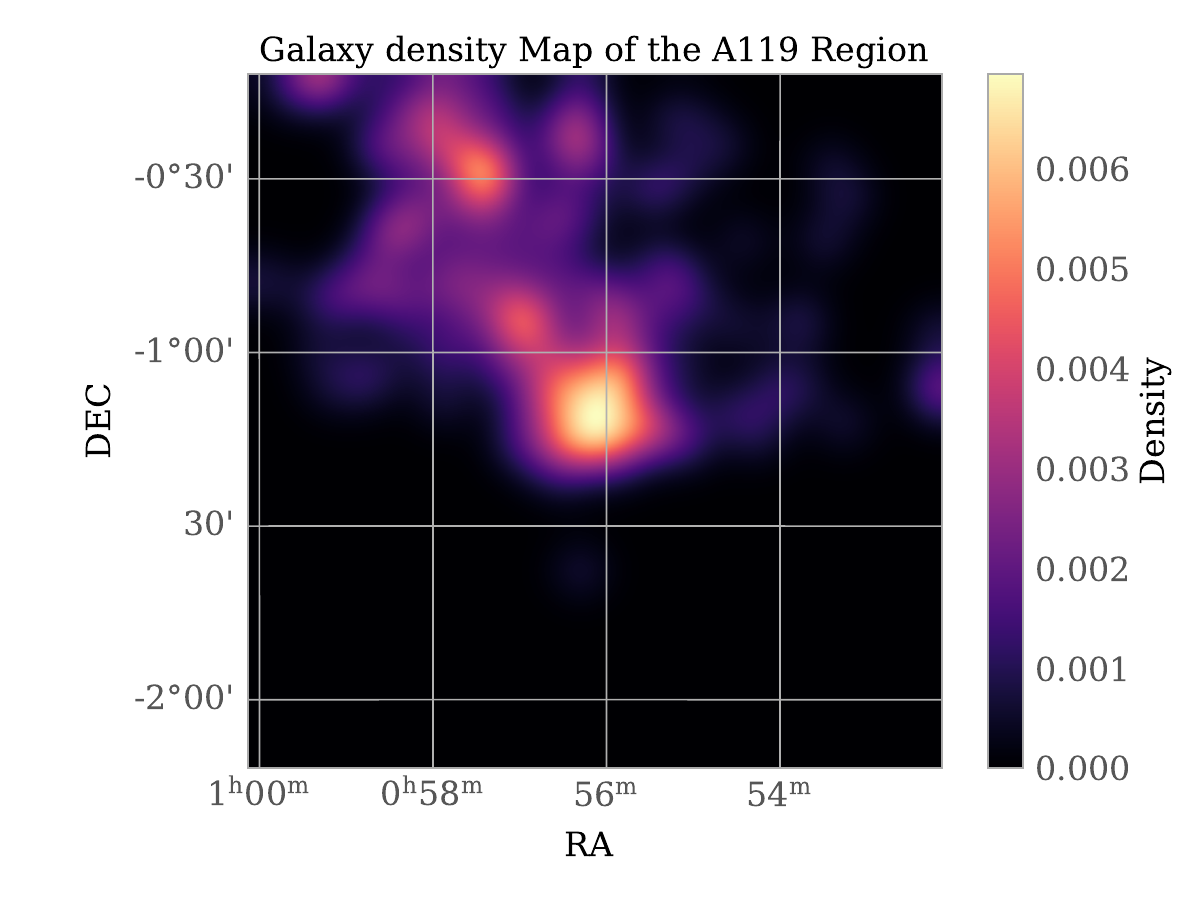}%
        }%
    \hfill%
    \subfloat[]{%
        \includegraphics[width=0.5\textwidth]{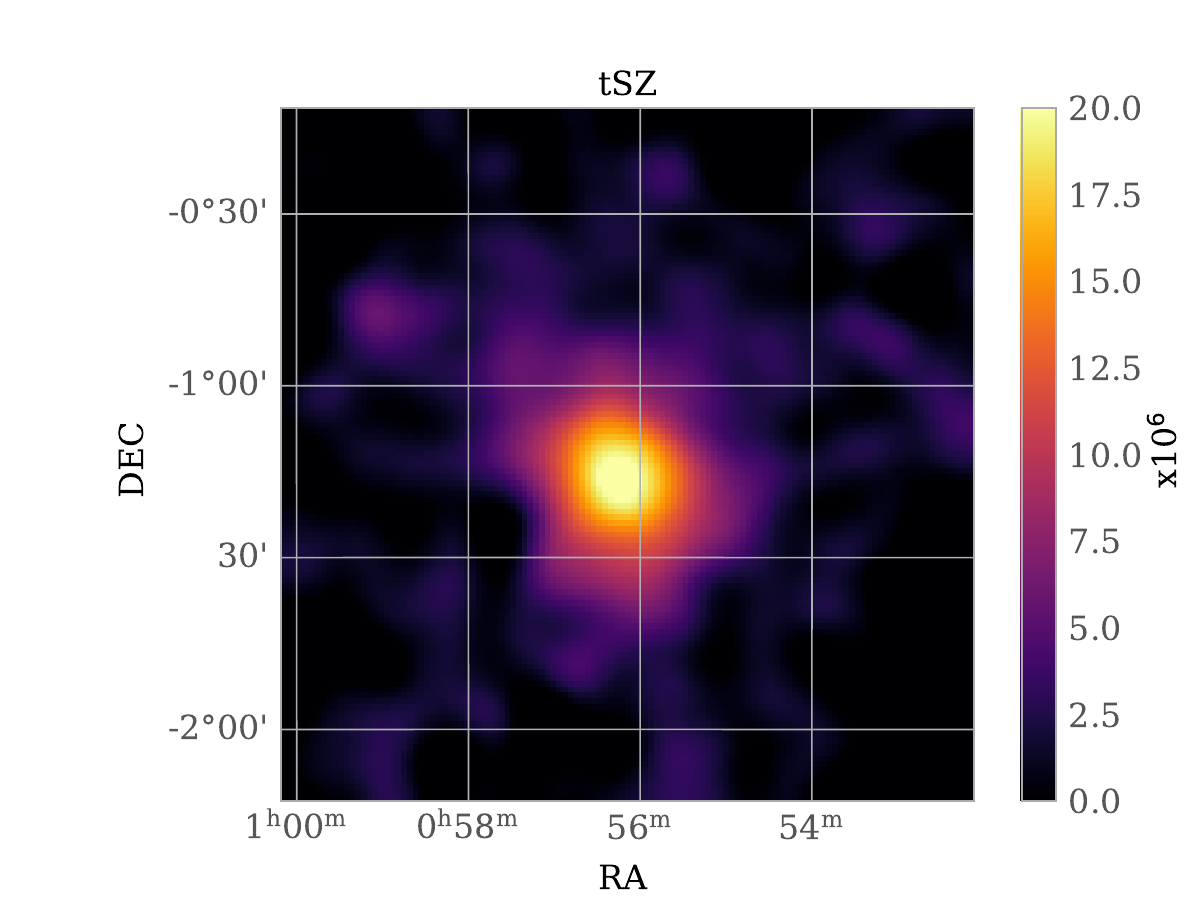}%
        }%
    \hfill%
    \subfloat[]{%
        \includegraphics[width=0.5\textwidth]{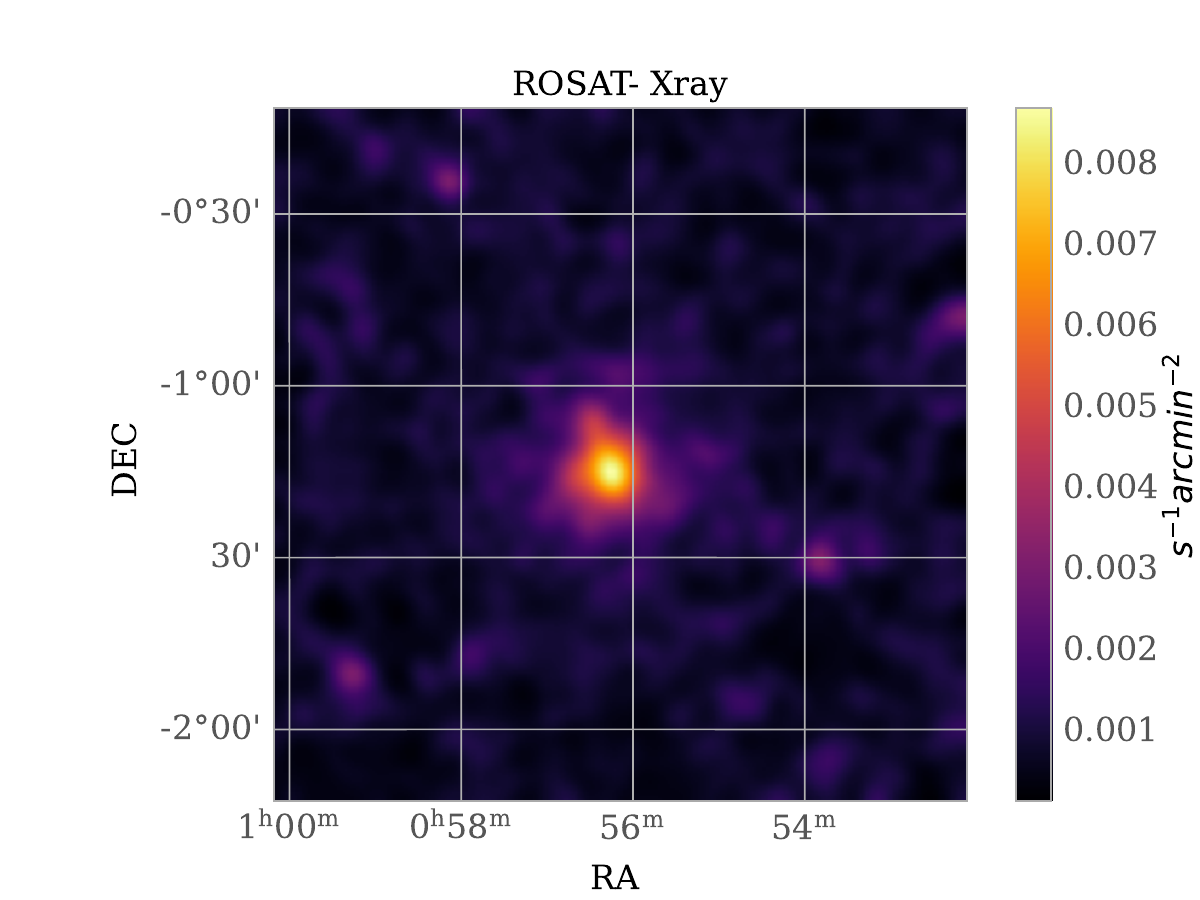}%
        }%
     \subfloat[]{%
        \includegraphics[width=0.5\textwidth]{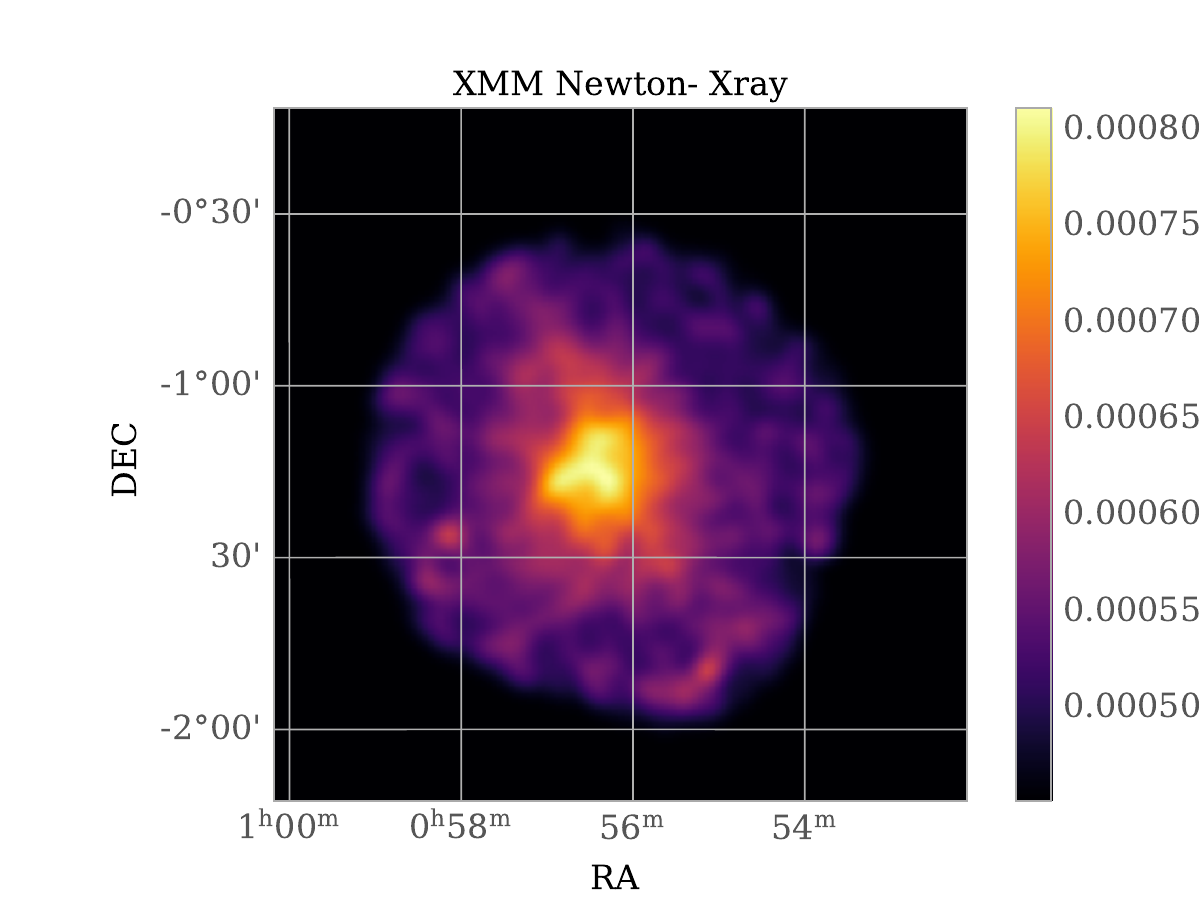}%
        }%
    
    \caption{Above images describe the $\gamma$-ray emission morphology within the galaxy cluster. From left to right and top to bottom: (1) density of galaxies, (2) thermal Sunyaev-Zel'dovich (tSZ) signal, (3) X-ray emission (ROSAT) , (4) X-ray emission (XMM Newton) ($s^{-1}arcmin^{-2}$). Each template provides a distinct observational perspective on cluster structure, aiding in the identification of features potentially correlated with $\gamma$-ray emission. }
    \label{fig:morphology check}  
\end{figure*}

To construct spatial templates shown in \autoref{fig:morphology check} from the multi-wavelength data, we considered templates based on galaxy density, X-ray surface brightness and the thermal Sunyaev-Zel'dovich (tSZ) effect. Each map is created on a $2\times2$ deg$^2$ except X-ray from XMM Newton grid with a pixel resolution of 1 arc-minute, ensuring a finer resolution than the \textit{Fermi}-LAT instrument. To reduce noise in these maps, a smoothing process was applied, though kept below \textit{Fermi}-LAT’s angular resolution to avoid impacting the map’s convolution with the instrument's response function. Additionally, to minimize noise-induced biases on larger scales, we mask out pixels more than 60 arcminutes from the reference center in the galaxy density, X-ray maps and tSZ maps. In each template, the spatial structure varies depending on its physical properties: the X-ray template, proportional to the square of the gas density, is the most compact. In contrast, the tSZ template, representing gas pressure, is more extended.

\subsection{Cosmic Ray proton (CRP) models }
These models focus on the energy distribution of the $\gamma$-ray signal and allow for detailed analysis of the emission spectrum. Spectral modeling provides insight into the processes driving the $\gamma$-ray production, such as interactions between cosmic-ray protons and intracluster gas or inverse Compton scattering. This approach offers a unique perspective for interpreting the data and elucidating the physical processes responsible for the observed emission.

The gas density profile of the intracluster medium (ICM) is crucial for modelling $\gamma$-ray emission in galaxy clusters because it serves as the target for proton-proton interactions, which produce $\gamma$-rays when cosmic-rays (CR) protons collide with thermal gas protons \citep{Pfrommer2004}. Additionally, the thermal pressure profile is essential as it provides a normalization factor for the amount of cosmic rays (CRs) in the cluster. This normalization links the distribution of thermal pressure to the expected cosmic ray pressure \citep{Reju2019MNRAS, Miniati2001}, helping to estimate the total CR content in the cluster and their contribution to the observed $\gamma$-ray emission. 

In this paper, we modelled the ICM of the cluster with electron density profile described by $\beta$-model \cite{Cavaliere1978} obtained from X-ray observations \cite{Sarazin1986}
\begin{equation}
     n_{gas}(r) = n_{gas}(r=0)\left[1+\left(\frac{r}{r_c}\right)^2\right]^{-\frac{3}{2}\beta}
 \end{equation}
 where $n_{gas}(r=0) = 1.18\times10^{-3}$cm$^{-3}$, $\beta=0.56$ and $r_c=378$ kpc is the core radius of A119 taken from \citep{Feretti_1999} rescaled with our cosmological model.

 The electron pressure profile follows the generalized NFW profile (gNFW) \citep{Nagai2007} model,
 \begin{equation}
     P_{gas}(r) = \frac{P_{gas}(r=0)}{x^{\gamma}(1+x^{\alpha})^{\frac{\beta-\gamma}{\alpha}}}
 \end{equation}
 where $x=r/r_s$, $r_s=R_{500}/C_{500}$, ($\alpha, \beta,\gamma$) are the slopes at the $r\sim r_s, r>>r_s$ and $ r<<r_s$ respectively. $R_{500}=1.194$~Mpc, $C_{500}=0.631$, $\alpha=1.4$, $\beta=5.5$, $\gamma<0.4$, $P_{gas}(r=0)=2.35e^{-3}$cm$^{-3}$~keV. Here we considered the universal model 'B' merger \citep[Morphologically disturbed; for more details, see][]{Arnaud2010,Planck_Collaboration2013}, and rescaled with our cosmological model.

\subsubsection{CRp models templates}

The diffuse $\gamma$-ray emission in the vicinity of the cluster is due to the cosmic ray proton interacting with the proton in the intra-cluster medium (ICM). The P-P interaction is responsible for the production of charged and neutral pions. Further, the neutral pions decay into $\gamma$-rays.
The cosmic ray proton density is a function of energy
\begin{equation}
    N_{CRp}(r,E)=X_{CRp}N(r,E)
\end{equation}
Where $X_{CRp}=\frac{U_{CRp}}{U_{th}}$ is the normalization factor with respect to the thermal energy component. 
\begin{equation}
    N(r,E)=N_1(r)N_2(E)
\end{equation}
Where $N_1(r)$ is the radial dependent component of cosmic ray proton. $N_1(r)$ can be expressed in terms of electron population or electron pressure as follows
\begin{equation}
    N_1(r)= A n_e^{\eta_{CRp}}(r) 
\end{equation}
\begin{equation}
    N_1(r)=AP_e^{\eta_{CRp}}
\end{equation}
$\eta_{CRp}$ will take possible values [$0,\frac{1}{2},1$] flat, extended and compact respectively \citep[see][for more details]{Adam2021}.
We have assumed $N_2(E)$ is the energy spectrum given by diffusive shock acceleration \citep[DSA;][]{ARBell1978,Blandford1978,Bladford1987} with a power law 
\begin{equation}
    N_2(E)=AE^{-\alpha_{CRp}}
\end{equation}
where $\alpha_{CRp}$ is the spectral slope. 

The cosmic ray proton (CRp) distribution in the cluster environment is characterized by three main parameters such as (I) Spatial Scaling Relative to Thermal Gas ($\eta_{CRp}$): This parameter defines how the spatial distribution of CRp scales in relation to the thermal gas distribution. It provides insight into how CRp are distributed within the cluster compared to the thermal gas. (II) CRp Spectrum Slope ($\alpha_{CRp}$): This parameter describes the spectral slope of the CRp energy distribution. The slope $\alpha_{CRp}$ is crucial for understanding the energy distribution and the nature of the CRp population, influencing the resulting $\gamma$-ray spectrum of CRp interactions with the intra-cluster medium. (III) Normalization factor ($X_{CRp}(R_{500})$): The normalization $X_{CRp}(R_{500})$ sets the absolute level of the CRp distribution at the characteristic radius $R_{500}$ provides a reference value for the CRp density within the cluster and is essential for scaling the CRp profile across different regions within the cluster. Together, these parameters define the spatial and spectral characteristics of the CRp distribution in the cluster, influencing the observable effects of the CRp interactions.

\begin{figure*}[ht] 
    \subfloat[]{%
        \includegraphics[width=0.5\textwidth]{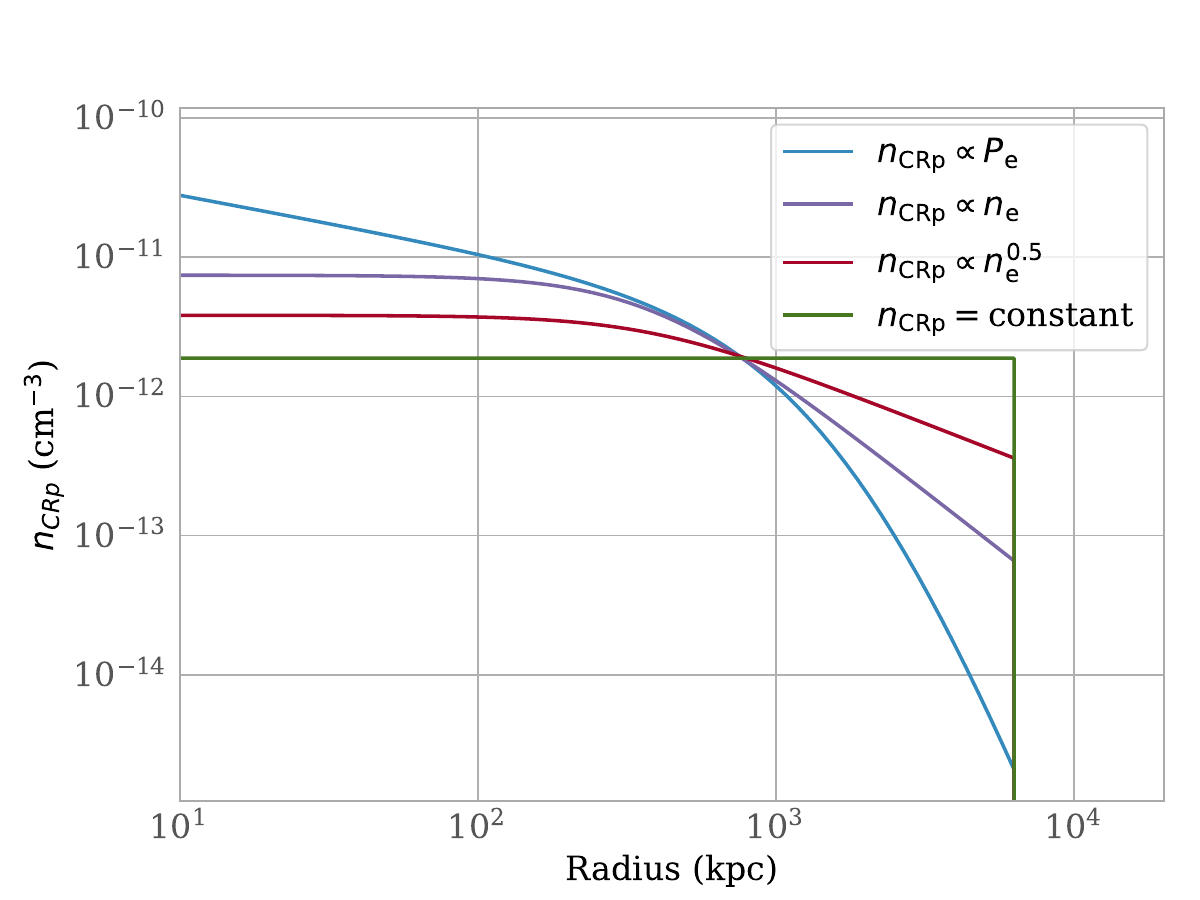}%
        }%
    \hfill%
    \subfloat[]{%
        \includegraphics[width=0.5\textwidth]{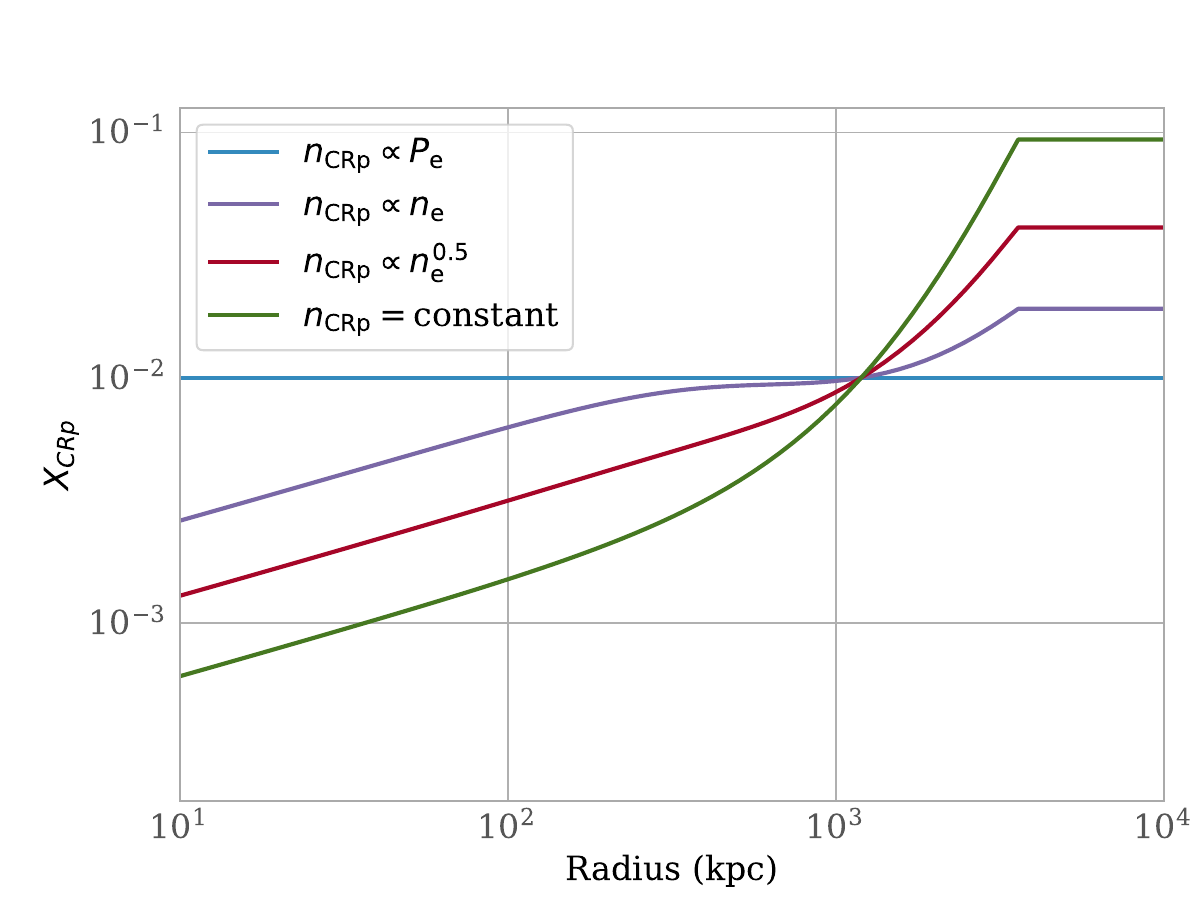}%
        }%
    \hfill%
    \subfloat[]{%
        \includegraphics[width=0.5\textwidth]{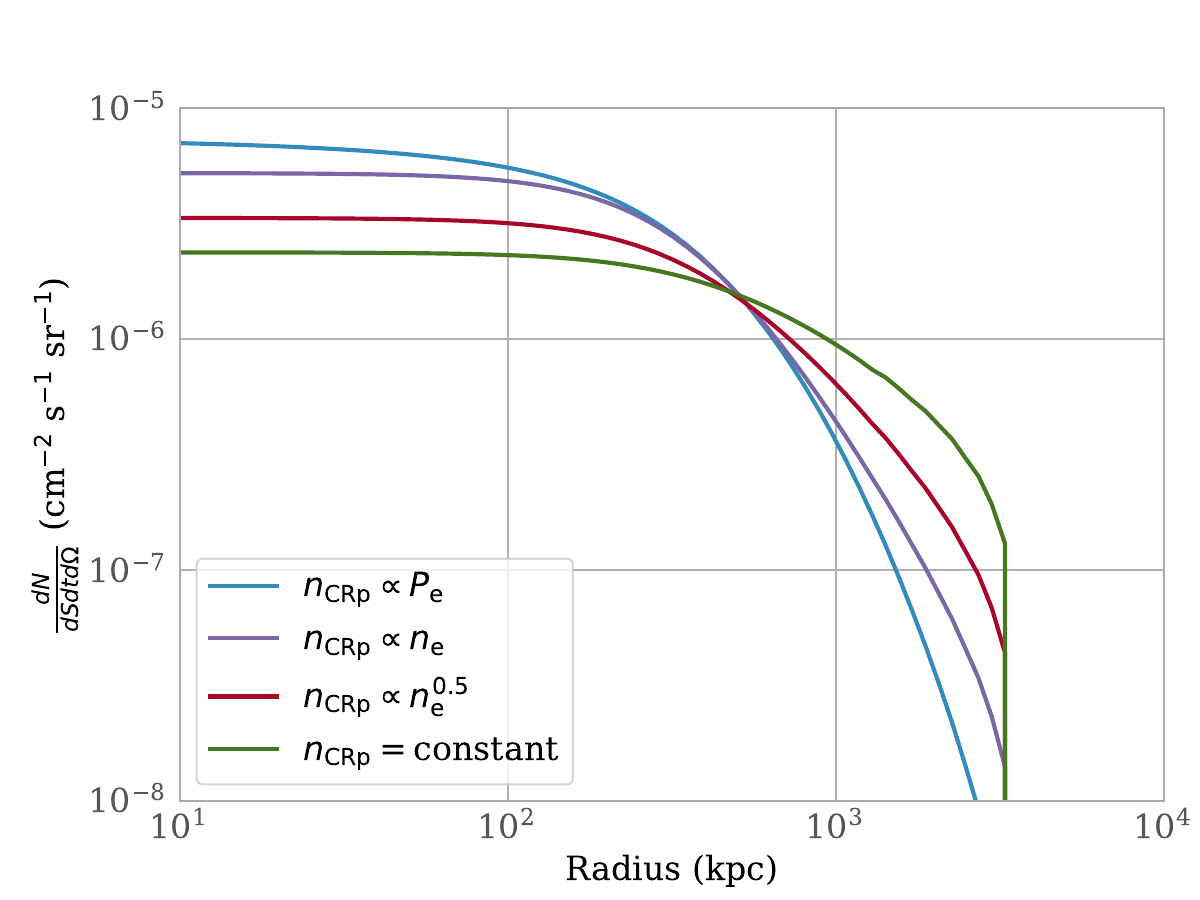}%
        }%
    \hfill%
     \subfloat[]{%
        \includegraphics[width=0.5\textwidth]{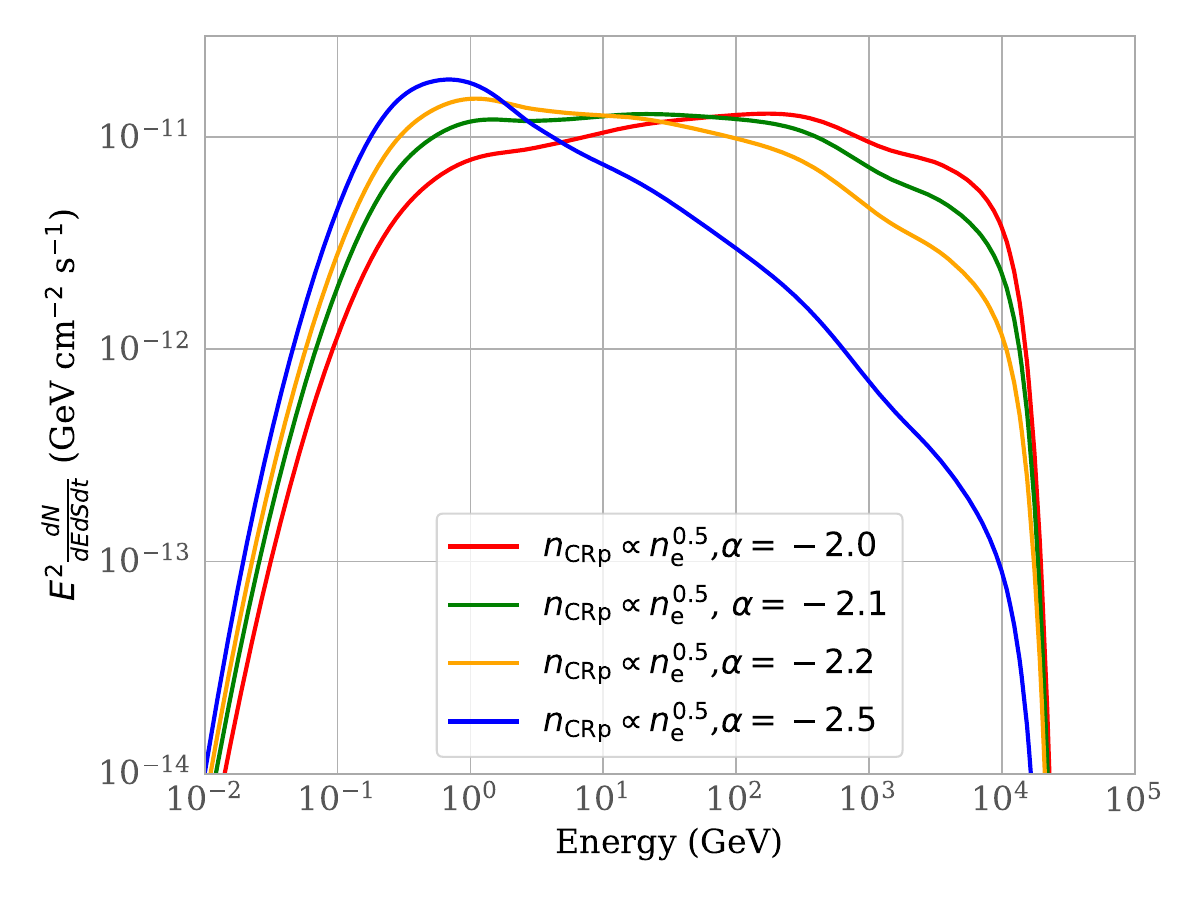}%
        }%
    \hfill%
    \caption{Comparison of MINOT model predictions for the cosmic-ray proton (CRp) distribution and associated $\gamma$-ray emission within a A119 cluster, under various assumptions about the CRp distribution. All models here are normalized to $X_{CRp}(R_{500})=10^{-2}$. Top left panel shows the radial profile of the CRp distribution (density). Top right panel shows ratio of enclosed CRp energy to thermal energy as function of radius. Bottom left panel shows the radial $\gamma$-ray surface brightness profile. Bottom right panel shows the integrated $\gamma$-ray spectrum within the $R_{500}$ }
    \label{fig:spectrum}  
\end{figure*}

We incorporated the $\gamma$-ray attenuation due to interactions with the extragalactic background light (EBL), which introduces a spectral cutoff above $E\gtrsim 10^4$ GeV. This attenuation, alongside spatial and spectral templates, is used to model the cluster in the Region of Interest (ROI). We have used the EBL model from MINOT \citep{MINOT2020} for our study that utilizes EBL model data from Flinke et al 2022 \citep{Finke2022ApJ}. In Figure (\ref{fig:spectrum}), we illustrated the impact of varying cosmic ray proton (CRp) profiles and spectral slopes on the $\gamma$-ray surface brightness and spectrum. The top left panel presents four different radial CRp distributions, which describe the CRp density as a function of radius. The top right panel shows the corresponding CRp to thermal energy ratio, with the normalization fixed at $X_{CRp}(R_{500})=10^{-2}$. The bottom left panel depicts the $\gamma$-ray surface brightness for the different CRp models, which is more concentrated than the CRp distributions due to the combined effects of CRp and thermal gas densities. The bottom right panel presents the $\gamma$-ray spectrum integrated within $R_{500}$ for different CRp slopes $\alpha_{CRp}$. The spectral shape between $1$ GeV and $10^4$ GeV is primarily driven by the CRp slope, while higher energies are truncated by the EBL cutoff, which is beyond the \textit{Fermi}-LAT range. At lower energies, the proton-proton collision threshold causes a smooth vanishing of the spectrum. The fixed normalization ensures that a steeper CRp spectrum results in a higher $\gamma$-ray flux near the peak. 
 
\begin{center}
\begin{table*}[t]%
\caption{ Test Statistic (TS) values and integrated flux in the 100 MeV−1 TeV energy band for different CRp distribution models applied to A119. Each model's TS value indicates the significance of the detection in each case, with higher values representing a better fit to the data within this energy range. The flux values provide the integrated $\gamma$-ray flux for each model, helping to compare predicted emission levels across different CRp assumptions.\label{tab1}}
\centering

\begin{tabular*}{500pt}{@{\extracolsep\fill}l c D{.}{.}{3}c@{\extracolsep\fill}}
\toprule

\toprule

\textbf{Model} & \textbf{TS} &  \textbf{Flux ($10^{-10}$ ph cm$^{-2}$ s$^{-1})$}  &  \multicolumn{1}{c}{\textbf{Loglike}}    \\
\midrule

Point Source & 8.39 &$8.71$\pm$2.94$  & -10032.114  \\
\midrule
\multicolumn{4}{c}{\textbf{2D Morphological Models}}  \\
\midrule 
Disc ($R=0.5^{\circ}$) & 17.98 &$11.48$\pm$3.00$  &  -10025.637  \\
Radial Gaussian & 17.94&$ 12.47$\pm$2.99$  & -10025.636  \\
Galaxy Density& 18.47  &$10.78$\pm$2.99 $  & -10025.650\\
X-ray (XMM)&  18.30&$11.64$\pm$2.89 $  &  -10025.593\\
Rossat X-ray& 18.32  &$10.62$\pm$2.86 $  & -10025.467 \\
tSZ& 18.04  &$12.10$\pm$3.00 $  & -10026.468 \\
\midrule
\multicolumn{4}{c}{\textbf{CRp Models}} \\
\midrule
Extended & 17.85 &$11.15$\pm$2.97$  &  -10024.147 \\
Isobar& 17.99  &$10.77$\pm$2.97 $  & -10025.333\\
Compact& 18.06  &$11.56$\pm$2.98 $  & -10025.272 \\
Flat& 17.43  &$10.52$\pm$2.94 $  & -10025.845 \\

\bottomrule \\ 
\end{tabular*}
\begin{tablenotes}

\item[1].  Extended model $n_{CRp}\propto n_{gas}^{\frac{1}{2}}$.
\item[2].  Isobar model $n_{CRp}\propto P_{gas}$.
\item[3].  Compact model $n_{CRp}\propto n_{gas}$.
\item[4].  Flat model $n_{CRp}=$ Constant.

\end{tablenotes}
\end{table*}
\end{center}

\begin{table*}
\caption{Comparison of spatial models relative to the disk template. The differences in Akaike Information Criterion ($\Delta$AIC) are computed with respect to the disk model ($\Delta$AIC = AIC$_i$ – AIC$_{\mathrm{disk}}$). Smaller $\Delta$AIC values indicate models that are statistically good model with respect to the disk in explanatory power.}
\label{tab:model_comparison}

\begin{tabularx}{\textwidth}{l *{5}{>{\centering\arraybackslash}X}}
\hline \\[0.5ex]

Model & log$L$ & $k$ & $\Delta$TS & AIC & $\Delta$AIC  \\[0.5ex] 
\hline\\ [0.5ex]
Gauss        & -10025.636 & 36 & 0.002  & 20123.271 & -0.003 \\[1ex]
Density      & -10025.650 & 36 & -0.026 & 20123.299 & 0.025    \\[1ex]
XMM          & -10025.593 & 36 & 0.088  & 20123.186 & -0.088  \\[1ex]
Rossat       & -10025.467 & 36 & 0.339  & 20122.934 &-0.340 \\[1ex]
Extended & -10024.147 & 35 & 3.0  & 20118.294 & -4.980                 \\[1ex]
Compact      & -10025.272 & 35 & 0.729  & 20120.545 & -2.729               \\[1ex]
Isobar    & -10025.333 & 35 & 0.607  & 20120.666 & -2.608                 \\[1ex]
Flat         & -10025.845 & 35 & -0.416 & 20121.690 &-1.584                 \\[1ex]
tSZ          & -10026.468 & 36 & -1.662 & 20124.935 & 1.661                                                 \\[1ex]
Point        & -10032.114 & 34 & -12.954& 20132.227 & 8.953              \\[1ex]
\hline
\end{tabularx}

\vspace{0.2cm}
\footnotesize{Models with $\Delta$AIC $\lesssim 0.05$ (Gauss, Density, XMM, Rossat) are essentially indistinguishable from the disk. The extended model shows the strongest relative preference ($\Delta$AIC $\approx -5$), suggesting a slight tendency toward extended emission. Compact, Isobar, and Flat yield moderate improvements ($\Delta$AIC$ \sim -1.5$ to $-3$), while the Point source is strongly disfavored ($\Delta$AIC \,$\gg 2$).}
\end{table*}

\section{ Data Analysis and Results}
Using 14 years of \textit{Fermi}-LAT Pass 8 data of A119 field, We first identified nearby 4FGL point sources—$4FGL J0059.3-0152$, $4FGL J0101.0-0059$, and $4FGL J0059.2+0006$—which exhibit significant TS values and are located within 1–2 degrees of the cluster center. For energies above 100 MeV, we observed a possible excess of diffuse $\gamma$-ray emission near A119, To investigate the nature of this excess diffuse $\gamma$-ray emission near and around the galaxy cluster A119, we computed Test Statistic (TS) values for various models fitted to the residual data. The models include point source, disc, radial Gaussian and CRp (radial) models, with results summarized in \autoref{tab1}, and the corresponding systematic uncertainties related to the extended models are reported in Appendix \ref{app11}.

\subsection{Localisation}

\begin{figure*}[h!]
 
    \subfloat[]{%
        \includegraphics[width=0.48\textwidth]{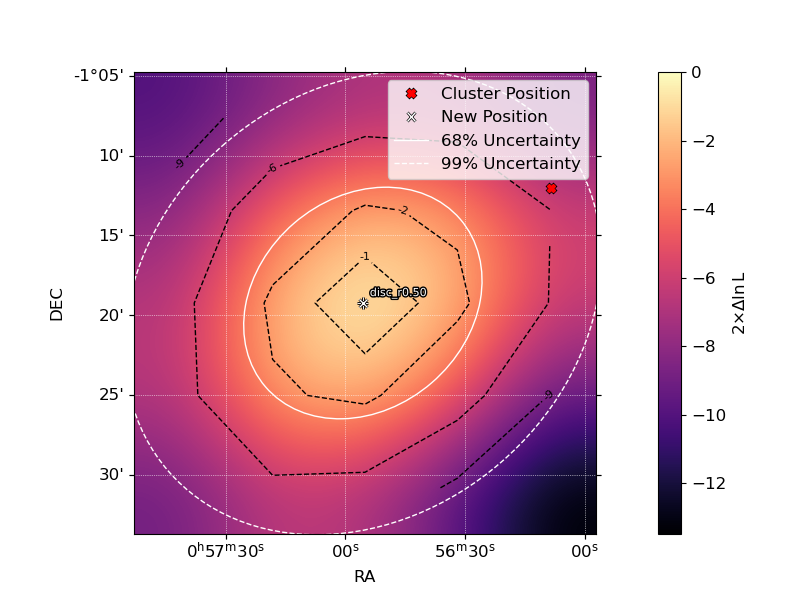}
        }%
    \hfill%
    \subfloat[]{%
        \includegraphics[width=0.48\textwidth]{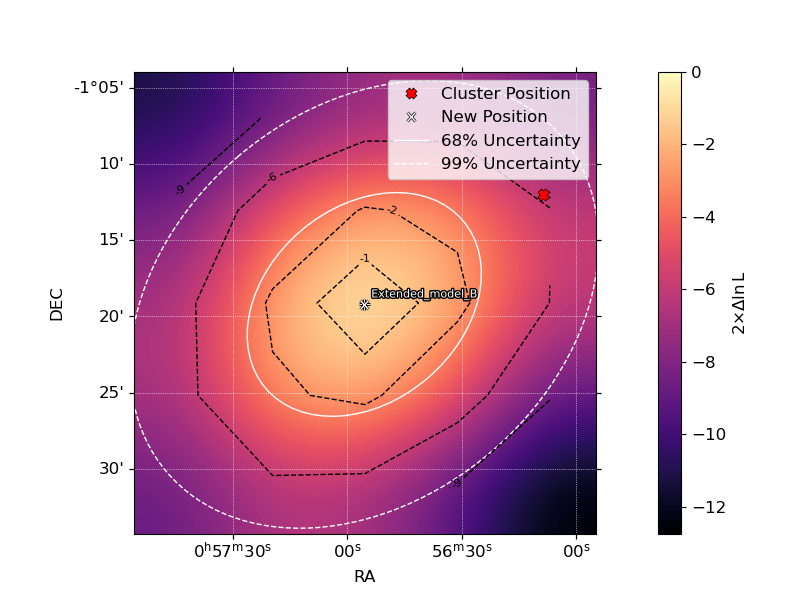}%
        }%
        \hfill%
 
    \caption{{\bf Left Panel:} shows disc model and {\bf Right panel:} shows extended model $(n_{CRp}\propto n_{gas}^{\frac{1}{2}})$: Localization map of the $\gamma$-ray source obtained with \texttt{Fermipy}. The color scale represents the likelihood surface in terms of $2 \times \Delta \ln L$, with warmer colors indicating regions of higher likelihood. The red cross marks the cluster position, while the black diamond denotes the best-fit position derived from the localization analysis. The solid and dashed contours correspond to the 68\% and 99\% confidence regions, respectively. For reference, the red cross indicates the center of the Abell~119 cluster providing spatial context for assessing the association between the localized $\gamma$-ray emission and the cluster environment.}
    \label{fig:localisation}
\end{figure*}

The source was localized using a likelihood-based analysis with function \texttt{localization} of \texttt{Fermipy} as shown in fig \ref{fig:localisation}. The best-fit coordinates are $\mathrm{RA} = 14.2319^\circ \pm 0.0827^\circ$ and $\mathrm{Dec} = -1.3202^\circ \pm  0.0792^\circ$, with an angular offset of $0.2300^\circ$ from the initial model position. The positional uncertainties are characterized by containment radii of $r_{68} = 0.1222^\circ$, $r_{95} = 0.1972^\circ$, and $r_{99} = 0.2446^\circ$, corresponding to the 68\%, 95\%, and 99\% confidence levels . Localization maps include the cluster center as a reference, providing spatial context for the source. The refined position lies within the vicinity of the cluster, indicating a likely association of the $\gamma$-ray emission with the cluster environment. These results demonstrate that the source is well-constrained.

\subsection{Morphological Analysis}\label{Morphological analysis}
The excess of diffuse $\gamma$-ray emission that we have identified in A119 cluster is located at RA$=14.0357^{\circ}$, DEC$=-1.20021^{\circ}$ (see the residual map in \autoref{fig:residual map}). For further analysis of this excess diffuse emission, we have categorized the \textit{Fermi}-LAT data into three bands viz $100$ MeV-$1$ TeV (Total energy band), $100$ MeV-$1$ GeV (middle energy band), and $1$ GeV-$1$ TeV (higher energy band). 
We modeled the signal with spatial point source morphology, 2D morphological models, and CRp models (radial models) such as extended model ($n_{CRp}\propto n_{gas}^{\frac{1}{2}}$), isobar model ($n_{CRp}\propto P_{gas}$), compact model ($n_{CRp}\propto n_{gas}$), and  flat model ( $n_{CRp}=$ constant). During fitting, within the ROI, the spectral parameter of the sources was kept free, the sources outside the ROI were kept fixed; the isotropic extragalactic emission model \textit{iso$\_$P8R3$\_$SOURCE$\_$V3$\_$v1.txt} and \textit{gll$\_$iem$\_$v07.fits} - the extragalactic diffuse background emission model were also made free.  We have removed the sources which have test statistic (TS;\citep{Neyman1928,Mattox1996})\footnote{$TS= -2ln(L_{max,0}/L_{max,1})$, where $L_{max,0}$ is the null hypothesis means only background and  $L_{max,1}$ source plus background} less than one after first iteration and repeated the same till all the sources have been removed from the sky model with TS value lower than one.  

Furthermore, we have used the \textit{find\_source} function of \textit{Fermipy} and to test the contamination from faint sources, we performed an iterative residual TS map search, adding sources at thresholds of TS $\geq 25, 16, 9, 2,$ and $1$ with re-optimization after each step. The cluster flux, TS, and morphology remained stable within uncertainties. The likelihood gains from very low-TS sources were negligible, confirming robustness against unresolved sources. Finally,  we reported sources with significance above $3\sigma$ only.

With the help of the \textit{tsmap} function of \textit{Fermipy}, we computed the TSmap for all the models. We have shown the residual excess map which has been computed using \textit{residue} function of \textit{Fermipy}, for the energy bands $100$ MeV - $1$ TeV, $100$ MeV-$1$ GeV and $1$ GeV-$1$ TeV shown in \autoref{fig:residual map}. \autoref{fig:residual map}, The left panel shows the residual excess map in the total energy band $100$ MeV-$1$ TeV. This remains similar in the residual excess map of the higher energy band $1$ GeV- $1$ TeV with a lower excess count shown in the extreme right panel of \autoref{fig:residual map}. We have used the \textit{localize} function of \textit{Fermipy}, to estimate the localized position of the above mentioned models. The localized positions of the aforementioned physical models are estimated and used as input for further analysis. We have also checked the source extension for each of the models using the \textit{extension} function of \textit{Fermipy}. For all the sources, we have allowed the normalization parameter to vary freely within the region of $4^{\circ}$ from the center of the ROI. The uniform disc model has been used to calculate the TS \value, and the same was repeated for other models as well. The TS value for each model is shown in \autoref{tab1}.

To verify the possibility of excess diffuse $\gamma$-ray emission due to point-like sources such as AGN or star-forming galaxies, radio galaxies \citep{Ackermann2016} in the cluster vicinity, 
we have used the \textit{find\_sources} function of \textit{Fermipy}. We found a point source at the localized position (RA, DEC $=14.58^{\circ},-0.35^{\circ}$ ) with TS$=10.88$.

The comparison of spatial templates, as summarized in the table (\ref{tab:model_comparison}), demonstrates that most models are statistically consistent with the disk model. The radial gaussian, Density, XMM, and Rossat models show $\Delta TS$ values close to zero. To assess the relative performance of the models, we also computed the Akaike Information Criterion (AIC \citep{Akaike1974}) for each case. AIC is defined as $ \mathrm{AIC} = 2k - 2\log L,$ where $( k )$ is the number of free parameters and $(log L )$ is the log-likelihood.  The $\Delta$AIC values are defined relative to the disk model as: $\Delta \mathrm{AIC} = \mathrm{AIC}_{\rm model} - \mathrm{AIC}_{\rm disk} $
and values computed to be $-0.003 $ , $0.025$, $-0.088$ and $-0.340$ for the radial Gaussian, Density, XMM, and Rossat models respectively, indicating similar explanatory power with minimal added complexity. The Extended model exhibits a largest improvement with $\Delta $AIC $\sim -5$, which corresponds to the strongest preference among the tested morphologies and suggesting the extended (ICM model) emission may provide a slightly better description of the excess $\gamma$-ray emission from the cluster. Other complex templates, such as Compact, Isobar, and Flat, provide moderate improvements in fit, as indicated by moderate negative ($-1.5\gtrsim \Delta$AIC$\gtrsim -3$) values which indicate some support but not at a statistically significant level. Models with larger $\Delta$AIC $>2$ values, such as the Point source model, indicate a strongly disfavored.

 \subsection{Null Hypothesis Test}
\begin{figure*}[ht] 
    \subfloat[]{%
        \includegraphics[width=0.48\textwidth]{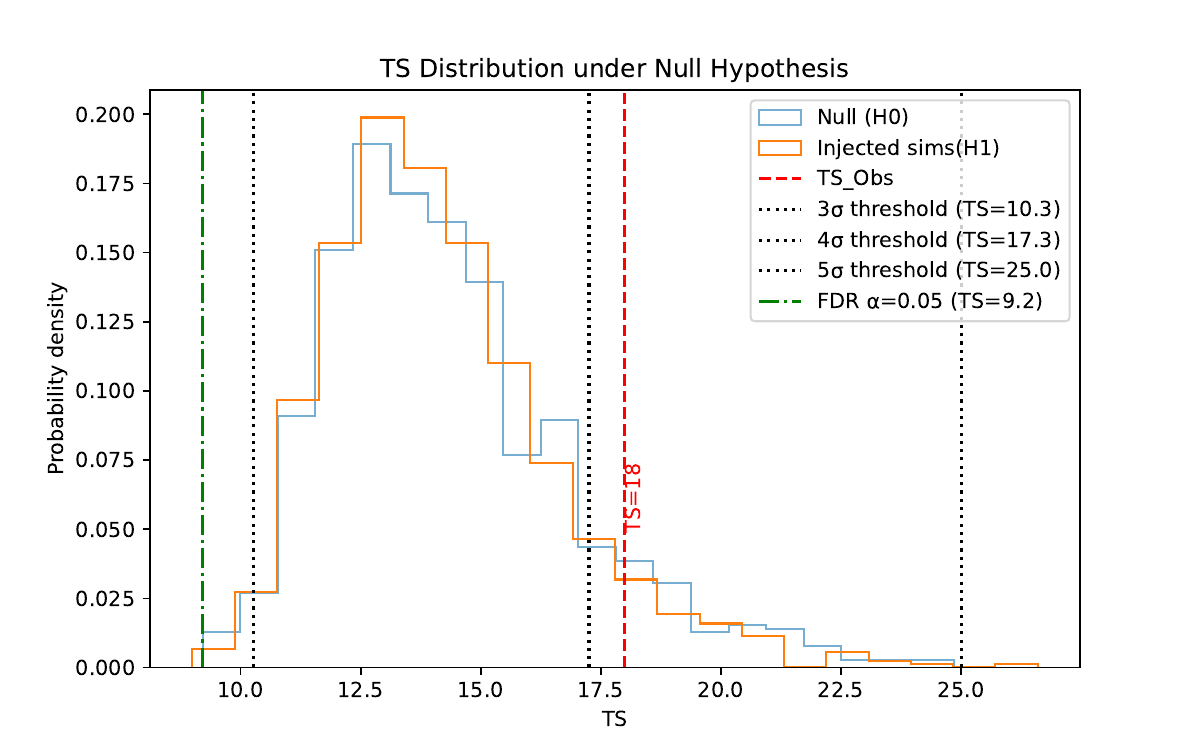}
        }%
    \hfill%
    \subfloat[]{%
        \includegraphics[width=0.48\textwidth]{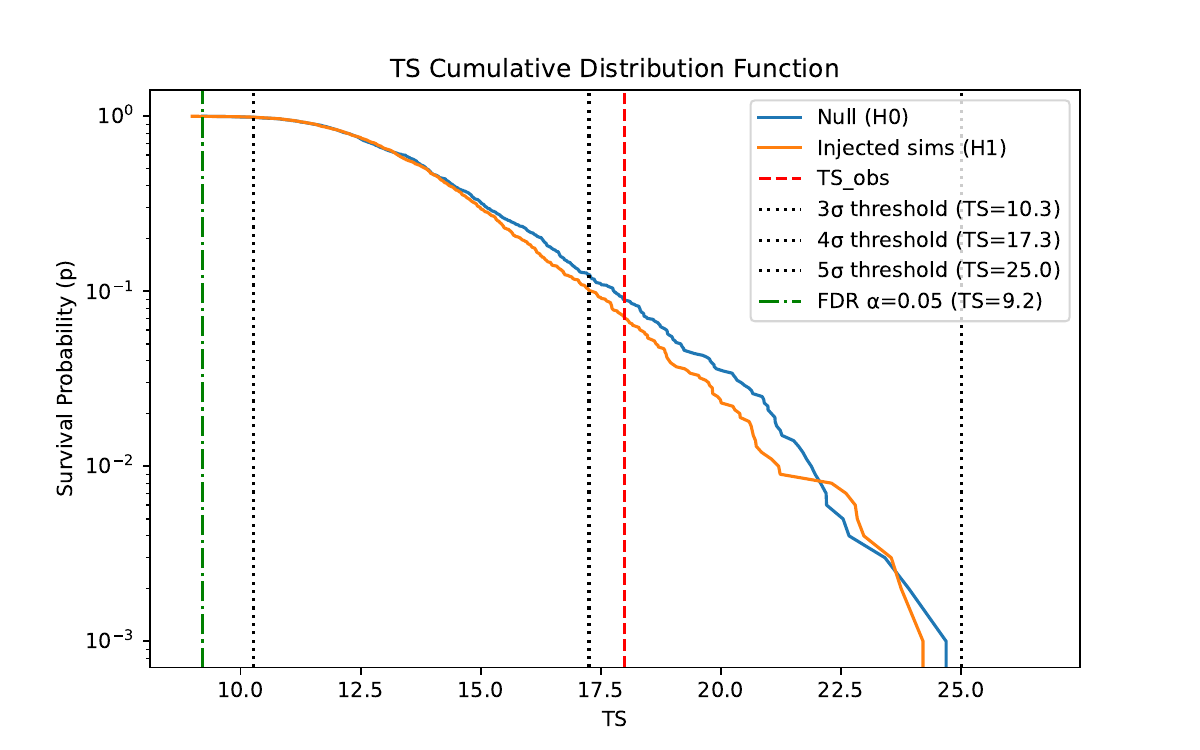}%
        }%
        \hfill%
    
    \caption{Left panel: Histogram of Test Statistic (TS) values from null simulations (blue) compared with injected source simulations (orange). The red vertical dashed line marks the observed TS ($\mathrm{TS_{obs}}=18$). Dotted lines denote empirical local significance thresholds corresponding to $3\sigma$ ($\mathrm{TS}\approx10.3$), $4\sigma$ ($\mathrm{TS}\approx17.3$), and $5\sigma$ ($\mathrm{TS}\approx25.0$). The green dash-dotted line indicates the FDR-corrected threshold ($\alpha=0.05$, $\mathrm{TS}\approx9.2$). Right panel: Cumulative distribution function (survival probability) of TS values under the background-only hypothesis (blue) and with an injected source (orange). The red vertical dashed line shows the observed TS, while dotted and dash-dotted lines again indicate the empirical local significance thresholds and the FDR-corrected threshold. }
    \label{fig:Null Hypothesis}   
 \end{figure*}

We evaluated the significance of $\gamma$-ray sources using a likelihood-ratio test under the null hypothesis of background-only counts. For detection, we first computed the local significance $\sigma_{\text{local}}$, which quantifies the probability that an observed excess at the fixed, pre-defined cluster position arises by chance. The resulting value of $\sigma_{\text{local}} \approx 4.1$ indicates a clear excess emission at the cluster location. To place these results in context, we further derived the global significance ($\sigma_{\text{global}}$) from null simulations covering the full region of interest, which accounts for the look-elsewhere effect. As expected, the $\sigma_{\text{global}}$ value is lower $(\sim1.35)$, reflecting the reduced probability once multiple search locations are considered. Finally, we applied a False Discovery Rate (FDR) correction to the local tests, under which the source formally passes, reinforcing that the excess seen locally is consistent across our analysis framework. Thus, while global corrections temper the overall significance, the local FDR-adjusted results highlight that, at the cluster position itself, we do observe a notable excess emission signal, as shown in the figure (\ref{fig:Null Hypothesis}).
 

\subsection{Analysis of Spectral Energy Distribution}
\begin{figure*}[ht] 
    \subfloat[]{%
        \includegraphics[width=0.5\textwidth]{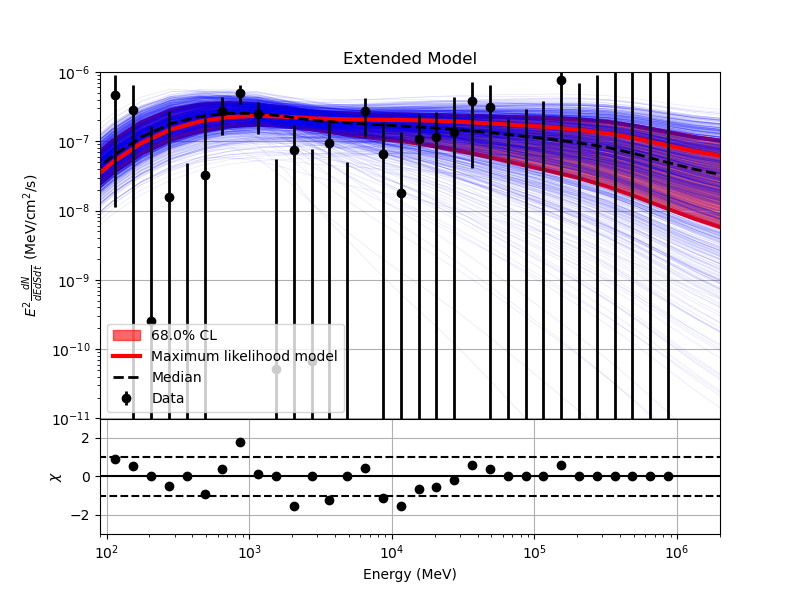}%
        }%
    \hfill%
    \subfloat[]{%
        \includegraphics[width=0.5\textwidth]{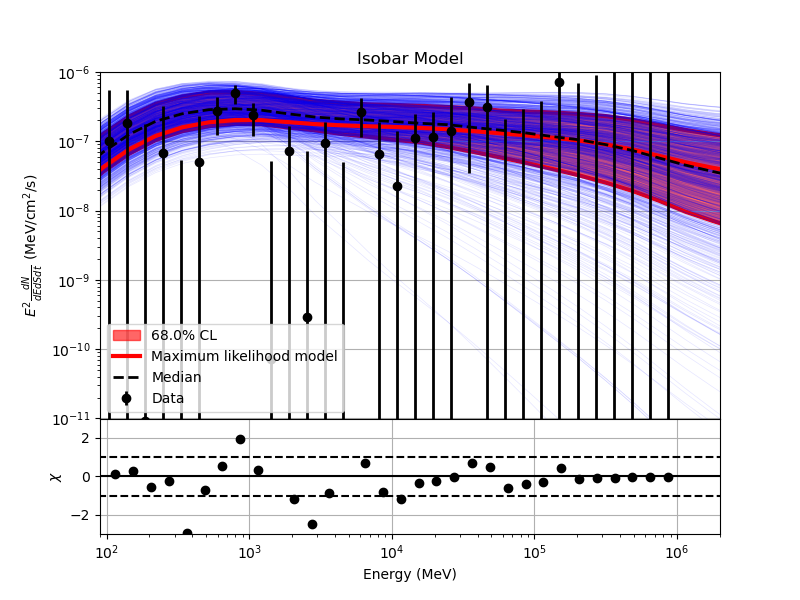}%
        }%
        \hfill%
    \subfloat[]{%
        \includegraphics[width=0.5\textwidth]{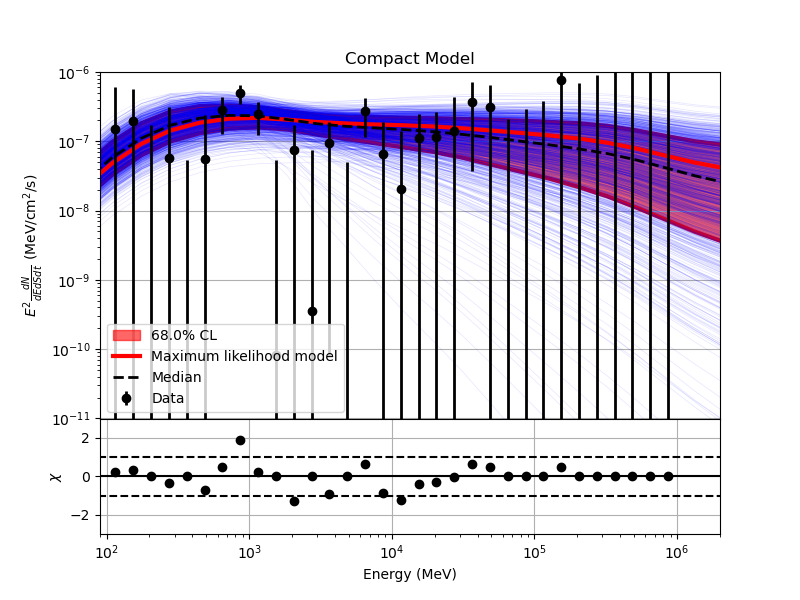}%
        }%
     \subfloat[]{%
        \includegraphics[width=0.5\textwidth]{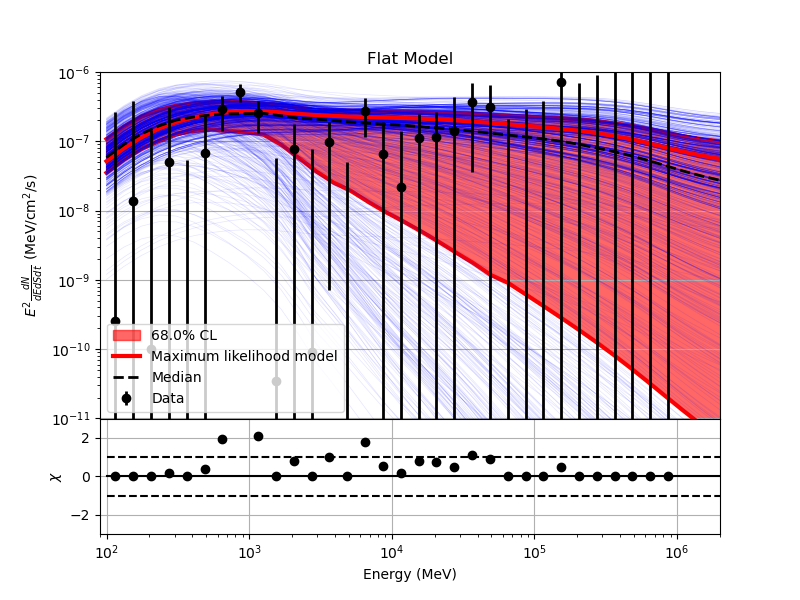}%
         }%
    \caption{For fixed background: The resultant Spectral Energy Distributions (SEDs) for A119, spanning an energy range from 100 MeV to 1 TeV, obtained with the \textit{SED} function of \textit{Fermipy}. From left to right and top to bottom: (1) Extended Model, (2) Isobar Model, (3) Compact Model, and (4) Flat Model, each fitted using MCMC. Each panel displays the best-fit SED for the respective model. The thick red line indicates the maximum likelihood model (best fitted model), and the light red region indicates the $68\%$ confidence level along with 1000 models obtained using MCMC simulation.The dash black line shows the median of these models. The residuals shown here are the difference between data and the best fit model, normalized with error-bars.}
    \label{fig:SED_with mcmc}   
 \end{figure*} 
 
The spectral energy distribution (SED) of the diffuse $\gamma$-ray emission from A119 was extracted using the \textit{sed} function in \textit{Fermipy}, where the photon slope allowed to vary as per the MINOT global spectral model. The analysis was performed under both fixed and free background scenarios. The SED method provides flux values and error bars for each energy bin, along with the likelihood scan in each bin for the normalization values. The integrated flux of the cluster diffuse emission between 100 MeV and 1 TeV shown in \autoref{tab1}, which ranges from $10$ to $12 \times 10^{-10}$ ph cm$^{-2}$ s$^{-1}$  for the CRp models. Similarly, the fluxes obtained for the 2D morphological models range from $11$ to $13 \times 10^{-10}$ ph cm$^{-2}$ s$^{-1}$.

Using the extracted $\gamma$-ray SED from fixed and free background scenarios, separately and CRp models, we constrain the cosmic ray (CR) population within the cluster. To constrain the CRp population, we use the \textit{Fermi}-LAT extracted SED, focusing on the hadronically induced $\gamma$-ray emission, which depends on the thermal gas pressure and density (from Planck and ROSAT data, respectively) and the CRp distribution, in this analysis, we kept thermal gas pressure and density as fixed parameters. With the fixed CRp spatial distribution, from the SED extraction, we vary only two parameters: (1) the CRp normalization, $X_{CRp}(R_{500})$, relative to thermal energy at radius $R_{500}$ and (2) $\alpha_{CRp}$, the CRp energy spectrum slope. These parameters are constrained across different models (extended, isobar, compact, flat) using the Markov chain Monte Carlo (MCMC) method using flat priors, $X_{CRp}(R_{500}) \in [0.0,0.2]$ and $\alpha_{CRp}\in [2.0,5.0]$. The likelihood function is evaluated per energy bin by interpolating the likelihood scan provided by \textit{Fermipy}, which maximizes parameter estimation efficiency. 

After the MCMC chains converge and the burn-in phase are removed, we integrated the two-dimensional histogram of $X_{CRp}(R_{500})$ and $\alpha_{CRp}$ to obtain constraints at a specified confidence interval. Marginalized posterior distributions provide individual parameter uncertainties, with $68\%$ confidence intervals defining the error bounds. Using these, we compute the model SED for each parameter set and create an envelope representing the $68\%$ confidence limit across all models in each energy bin. This method also determines the $\gamma$-ray flux and luminosity between 100 MeV and 1 TeV based on the MCMC sampling results shown in \autoref{tab2}.

The SEDs for each model are illustrated in \autoref{fig:SED_with mcmc} and \autoref{fig:SED_with mcmc_free_background}, corresponding to the fixed and free background scenarios, respectively. Error bars denote the $1\sigma$ uncertainty, derived from the curvature of the likelihood scan. The maximum likelihood model is highlighted in red, accompanied by 1,000 samples drawn from the MCMC chains, with their median shown as a dotted black line and the $68\%$ confidence interval shaded in red. The extended model demonstrates the best alignment with the observed data across both fixed and free background scenarios, indicating its robustness. The isobar model provides a slightly weaker fit, but still aligns reasonably well in both cases. In contrast, the compact and flat models show moderate agreement under the fixed background scenario but perform poorly when the background is allowed to vary freely, as confirmed by the residuals analysis. The models exhibit strong constraints at lower and intermediate energy bands, where uncertainties are minimal. However, at higher energies, uncertainties grow significantly, with the spectral amplitude remaining almost unchanged and error contours broadening, reflecting the limitations in constraining the spectrum at these energies.

\begin{figure*}[ht] 
    \subfloat[]{%
        \includegraphics[width=0.45\textwidth]{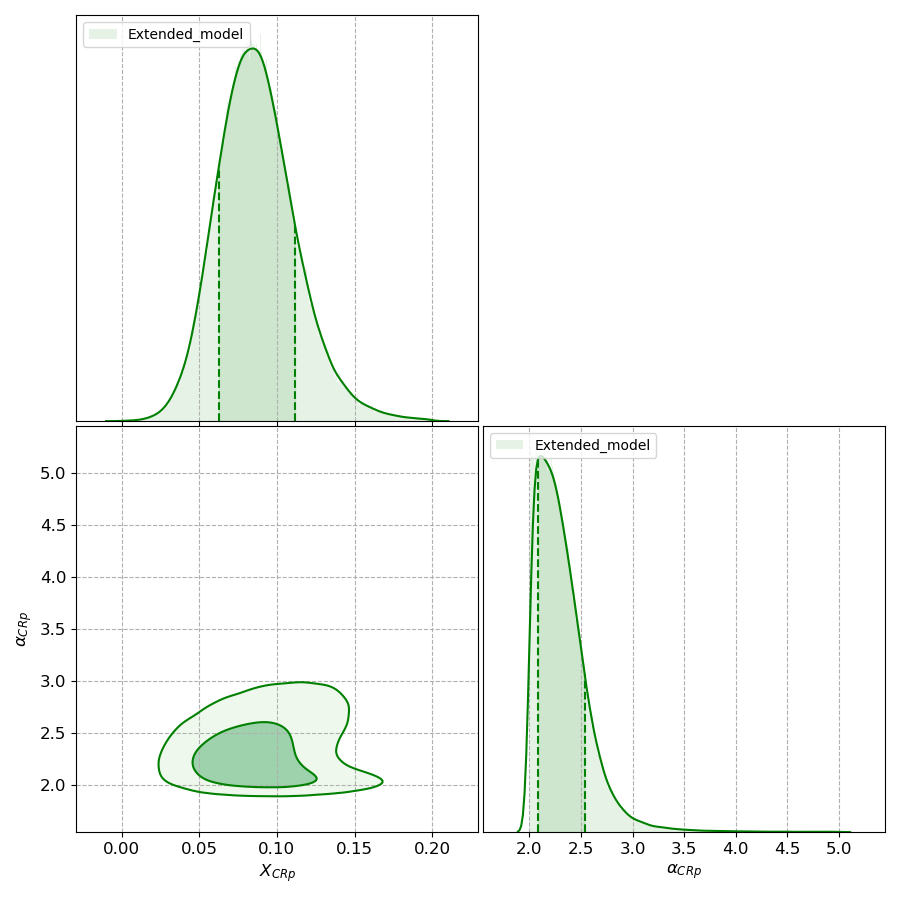}%
        }%
    \subfloat[]{%
        \includegraphics[width=0.45\textwidth]{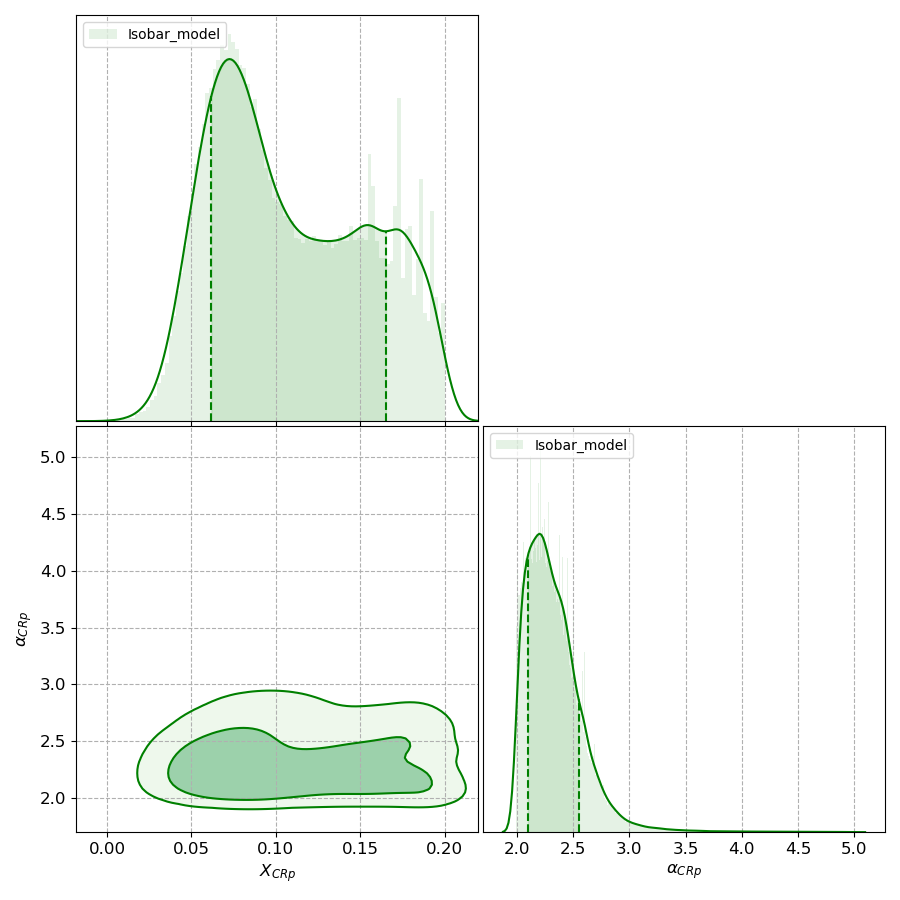}%
        }%
         \hfill%
    \subfloat[]{%
        \includegraphics[width=0.45\textwidth]{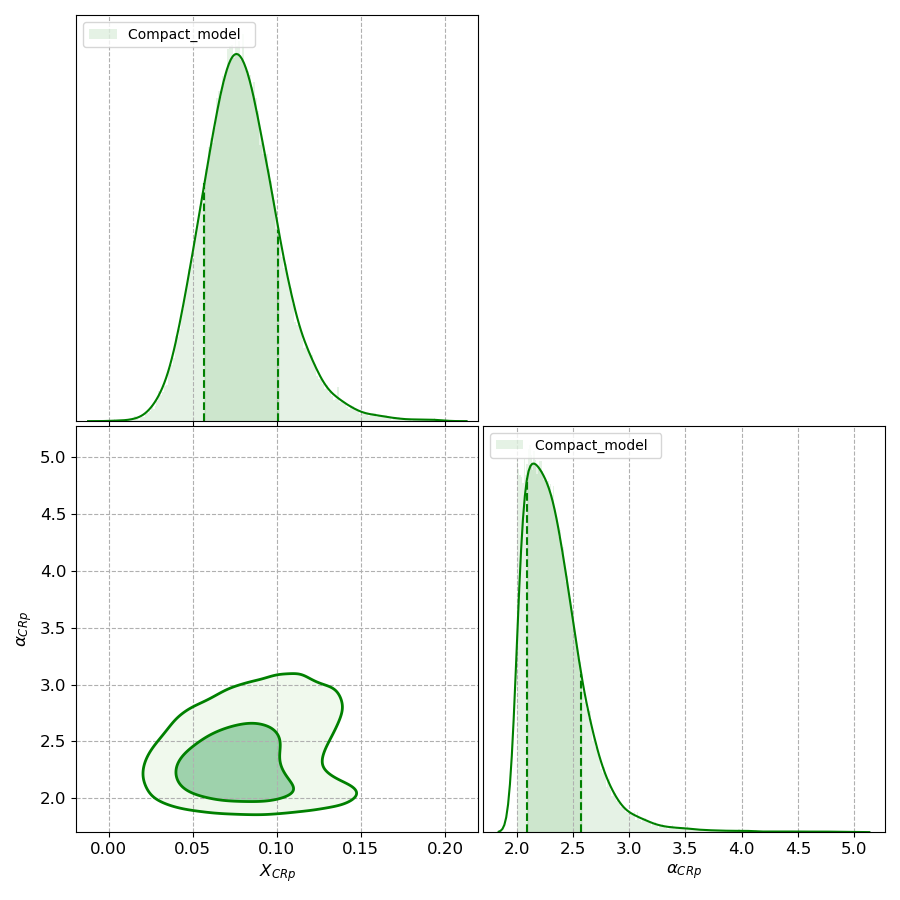}%
        }%
    \centering       
     \subfloat[]{%
        \includegraphics[width=0.45\textwidth]{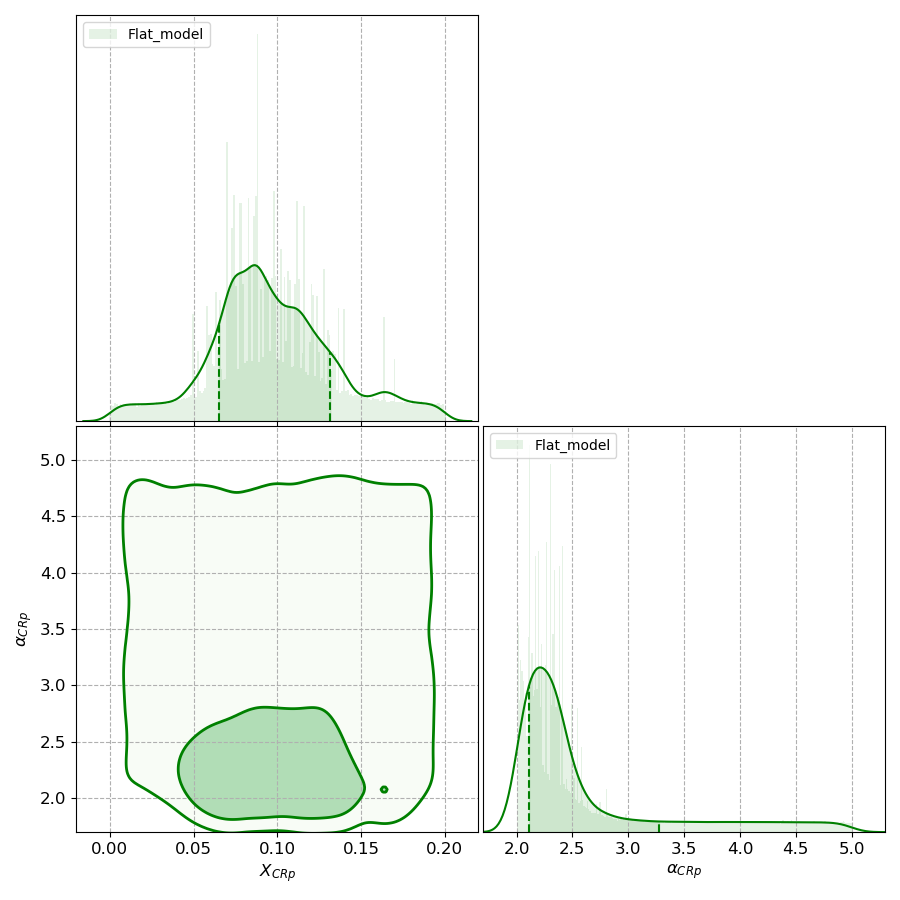}%
         }%
             
        \caption{Fixed Background: MCMC models constraints, displayed as corner plots. From left to right and top to bottom: (1) Extended Model, (2) Isobar Model, (3) Compact Model, and (4) Flat Model. Each plot presents the posterior distributions and parameter correlations, reflecting the MCMC sampling results for each model scenario. The posterior probability distribution is shown at the top for the $X_{CRp}(R_{500})$ and at the bottom right for the $\alpha_{CRp}$, where the green shaded area indicates the $68\%$ confidence interval. The bottom left panel shows the contour levels at the $68\%$ confidence interval in the dark green shaded area and the $95\%$ confidence interval in the light green shaded area for the $X_{CRp}(R_{500}) - \alpha_{CRp}$.}
    \label{fig:corner}    
 \end{figure*}

The posterior likelihood constraints for $X_{CRp}(R_{500})$ and $\alpha_{CRp}$ of each model are shown in \autoref{fig:corner} with fixed background and \autoref{fig:corner_freebackground} with free background. We present the MCMC constraints on $X_{CRp}(R_{500})$ and $\alpha_{CRp}$ for all tested spatial CRp models. Additionally, the fluxes and associated luminosities are derived and rigorously constrained using the MCMC best-fit models. (1) Extended model: The constraints on the SED models result in tight bounds on both parameters, $X_{CRp}(R_{500})$ and $\alpha_{CRp}$  in both fixed and free background scenarios, as illustrated in the top-right panels of \autoref{fig:corner} and \autoref{fig:corner_freebackground}, respectively. In the fixed background scenario, the CRp-to-thermal energy ratio $X_{CRp}(R_{500})$  is constrained to approximately $8\%$ and slope $\alpha_{CRp}$ constrained to approximately $2.18$.  In the free background scenario, $X_{CRp}(R_{500})$ is constrained to approximately $7.7\%$ and $\alpha_{CRp}$ approximately $2.25$, as shown in the marginalized distributions. (2) Isobar model: The constraints on the SED models reveal a tight bound on $\alpha_{CRp}$ and a relatively loose bound on $X_{CRp}(R_{500})$  in the fixed background scenario, as illustrated in the top-left panel of \autoref{fig:corner}.  In the free background scenario, the bounds are reversed, showing a fairly loose constraint on $\alpha_{CRp}$ and tight bound on $X_{CRp}(R_{500})$ as shown in top-left panel of \autoref{fig:corner_freebackground}.(3) Compact model: The constraints on the SED models result in tight bounds on both parameters, $X_{CRp}(R_{500})$ and $\alpha_{CRp}$ in the fixed background scenario, as illustrated in the bottom-right panel of \autoref{fig:corner}. In the free background scenario, however, no meaningful constraints are obtained for either parameter, as shown in the bottom-right panel of \autoref{fig:corner_freebackground}.(4) Flat model:In both the fixed background and free background scenarios, no meaningful constraints are found for either parameter, as shown in the bottom-left panels of \autoref{fig:corner} and \autoref{fig:corner_freebackground}, respectively. 

\autoref{tab2} and \autoref{tab3_freebackground} presents the MCMC constraints on $X_{CRp}(R_{500})$ and $\alpha_{CRp}$ across all tested spatial models. Additionally, the table includes the fluxes and corresponding luminosities, as derived from the MCMC fits for each model, providing a comprehensive view of the CRp population's impact on the A119, diffuse $\gamma$-ray emission.
\begin{table*}[ht!]
\caption{Constraints on the CRp Population, Flux, and Luminosity for CRp Models — This table presents the MCMC-derived constraints on the CRp normalization $X_{CRp}(R_{500})$ and spectral slope $\alpha_{CRp}$ along with the associated $\gamma$-ray flux and luminosity values, for each tested CRp model configuration. \label{tab2}}
    \begin{tabular}{ccccc}
    \toprule \\
ICM or Sky Model       &  $X_{CRp} $     &  $\alpha_{CRp}$  & Flux ($10^{-10}$ ph cm$^{-2}$ s$^{-1}$ )  & Luminosity ($10^{42}$ erg s$^{-1}$) \\ \\
\midrule 
Extended Model($n_{CR_P}\propto n^{1/2}_{gas}$)&  $0.08^{+0.04}_{-0.01}$  &$2.18^{+0.35}_{-0.09}$  &$12.36^{+9.26}_{-3.24}$  & $12.21^{+2.74}_{-3.95}$  \\ \\
Isobar Model($n_{CR_P}=P_{gas}$)  &$0.07^{+0.10}_{-0.00}$ &$2.22^{+0.33}_{-0.12}$  &$11.06^{+19.28}_{-1.19}$  &$9.80^{+11.54}_{-1.75}$   \\ \\
Compact Model($n_{CR_P}\propto n_{gas}$)  & $0.07^{+0.32}_{-0.01}$   & $2.22^{+0.35}_{-0.13}$ & $11.83^{+7.98}_{-3.58}$ &$10.45^{+2.92}_{-3.18}$   \\ \\
Flat Model($n_{CR_P}=$Constant) & $0.09^{+0.04}_{-0.02}$  &$2.22^{+0.46}_{-0.11}$  &$7.18^{+4.47}_{-2.25}$  & $6.46^{+1.63}_{-2.54}$ \\ \\

\bottomrule 
\end{tabular}
\begin{tablenotes}

\item \textbf{Note:} The reported values and uncertainties represent the maximum likelihood estimates along with the $68\%$ confidence intervals derived from the distributions.

\end{tablenotes}
\end{table*}

\section{Discussion}\label{sec3}

\subsection{Diffuse $\gamma$-Ray Emission from A119}
 
We performed the analysis of 14 years of LAT Pass8 data above $100$ MeV for the A119 cluster, in which we observed an excess of diffuse $\gamma$-ray emission in the vicinity of the A119 with $4.34\sigma$ confidence, as shown in \autoref{fig:residual map}. 
The residual map in \autoref{fig:residual map} reveals an interesting fact: the residual is the background-subtracted emission from the cluster region, partly overlapping with the virial radius of the cluster A119. This makes our analysis comparable to the emission predicted from similar sources in the vicinity of the cluster.

The best-fit computed integral flux for the extended (spectral) models  with spectral slopes of 2.18 to 2.25, relates to the measured spectral slopes in \citep{Hussain2023} ,\citep{Nishiwaki2021}. For the extended model, the computed luminosity bounds are $12.21_{-3.33}^{+2.74}\times10^{42}$ erg s$^{-1}$ in the fixed background scenario and  $11.49_{-3.33}^{+3.52}\times10^{42}$ erg/s in the free background scenario, indicating consistent results across both cases despite differences in background treatment. 

\subsection{Diffuse Emission from ICM}
The diffuse ICM $\gamma$-ray emission can only be quantitatively computed for specific models for the ICM energetic protons responsible for this emission through interactions with ICM gas via  $\pi^{0}$ decay. A comprehensive model must consider the particle source distribution, propagation mode, spatial distribution of ICM gas, magnetic field, and non-thermal particles \citep{Brunetti2014}. Our analysis considers only the particle source distribution and spatial distribution of ICM gas.

The estimated $\gamma$-ray emission in our model is approximately $9.26\pm2.81\times10^{-10}$ ph cm$^{-2}$ s$^{-1}$, with integrated flux values obtained through MCMC fitting of $12.36^{+9.26}_{-3.26}\times10^{-10}$ ph cm$^{-2}$ s$^{-1}$ in the fixed background scenario and $13.94^{+10.26}_{-4.43}\times10^{-10}$ ph cm$^{-2}$ s$^{-1}$ in the free background scenario. These flux values align well with the expectations of extended $\pi^0$-decay $\gamma$-ray emission as shown in \autoref{fig:SED_with mcmc}. A summary of the minimal predicted $\gamma$-ray flux in the direction of the A119 cluster is provided in \autoref{tab1}. The residual emission identified in this analysis suggests the possible presence of diffuse $\gamma$-ray emission in the region of the A119 cluster. This residual signal, though not yet conclusively detected, may become more pronounced with additional observations and in future with the improvement in sensitivity and resolution of the telescopes. Such diffuse emission can be interpreted within the framework of the hadronic emission mechanism, wherein cosmic-ray protons interact with the intracluster medium to produce $\pi^0$-decay into $\gamma$-rays. This mechanism offers a plausible explanation for the detected residual emission and its spatial correlation with the cluster region.

This observation strongly indicates that the $\gamma$-ray emission detected by Fermi-LAT can be predominantly attributed to hadronic processes. While this provides a plausible explanation for the observed $\gamma$-ray signal, to further probe this scenario, we estimated the corresponding neutrino flux upper-limit as $E^{2}\phi_{\nu} \approx 3 \times 10^{-10} \ \mathrm{GeV~cm^{-2} s^{-1} sr^{-1}}$. This value remains below the current detection threshold, so can provide this additional support for the proposed emission mechanism only in the future \citep{Zhang2025APh}.

Most spatial templates are statistically consistent with the disk model, exhibiting minimal differences in explanatory power (for detail see in section \ref{Morphological analysis} ). Among the tested morphologies, the Extended model shows the strongest preference, suggesting a slight enhancement in capturing more complex $\gamma$-ray structures, while other CRp models offer modest improvements. In contrast, the Point source model is clearly disfavored, confirming that the emission is predominantly ICM rather than point-like.

We also evaluated systematic uncertainties in the spectral parameters of the extended models due to the IEM, event class, low-energy threshold, and PSF0–3 selection, results summerised in Table \ref{tab:systematics effects}. Flux uncertainties are below 20\% for the IEM, 15\% for event class, $<$30\% for raising the threshold from 100 MeV to 500 MeV, and ~37\% for PSF0–3, consistent within the calculated error-bars. Overall, the dominant effects arise from the diffuse background and PSF selection, but the flux and spectral trends remain robust.

\section{Conclusions}\label{sec5}

Our analysis of the 14-year \textit{Fermi}-LAT data for the A119 cluster region reveals a potential excess of diffuse $\gamma$-ray emission. This residual emission significantly overlaps with the cluster's virial radius and halo of emission in other wave bands (e.g., X-ray, SZ etc.), hinting at a potential association with the cluster's environment. The Extended model reveals a subtle but discernible preference for more intricate $\gamma$-ray emission structures. This suggests that accounting for mildly complex morphologies can provide a more accurate characterization of the cluster’s high-energy emission. To further investigate this emission, we employed the \textit{Fermipy} software to generate Test Statistic (TS) maps and conducted a comprehensive spatial and spectral analysis. The TS maps, overlaid on the residual maps, highlighted regions of potential $\gamma$-ray emission, providing valuable insights into the spatial distribution of the excess emission. The spectral energy distribution (SED) was extracted using \textit{Fermipy's} \texttt{sed} function and subsequently analyzed using Markov Chain Monte Carlo (MCMC) techniques. The extended model provided the best fit to the data, yielding constraints on cosmic ray protons (CRp) relative to the thermal energy in the cluster. In the fixed background scenario, the CRp-to-thermal energy ratio was found to be $X_{CRp}=\frac{U_{CRp}}{U_{th}} \approx 8\%$ with a spectral slope of $\alpha_{CRp}=2.18^{+0.35}_{-0.09}$. In the free background scenario, the ratio was slightly lower, at $X_{CRp}=\frac{U_{CRp}}{U_{th}} \approx 7.7\%$ with spectral slope $\alpha_{CRp}=2.25^{+0.38}_{-0.13}$. The computed luminosity bounds for the extended models are consistent with previous studies 
and provide further evidence for the presence of high-energy phenomena in galaxy clusters.
Although, the above information are indicative of hadronic origin of $\gamma$-rays from the ICM of this cluster, the detection of the estimated neutrino flux $E^{2}\phi_{\nu} \approx 3 \times 10^{-10} \ \mathrm{GeV  ~cm^{-2} s^{-1} sr^{-1}}$ is essential for further confirmation. This definitely provides motivation for observation with sensitive neutrino telescopes such as IceCube-Gen2 \citep{Aartsen2021}, The Giant Radio Array for Neutrino Detection (GRAND) \citep{Alvarez-Mu2025}, KM3NeT (Cubic Kilometre Neutrino Telescope) \citep{KM3NeTCollaboration2025Natur} etc, in the future in the high energy range.

While the observed greater than $4\sigma$ signal significant, yet we cannot confirm a detection, mainly due to poor source localization, uncertainties and other instrumental limitations of the current telescopes. Nevertheless, it highlights the potential for deeper exploration of non-thermal processes in the cluster environment with improved sensitivity and angular resolution in future. Furthermore, an excess of $\gamma$-rays from galaxy clusters has far reaching implications in resolving $\gamma$-ray background problem and this result is significant in that respect.

\section*{Acknowledgments}
We are grateful to the \textit{Fermi}-LAT Collaboration for providing the data and analysis tools. Special thanks to the developers of the \textit{Fermipy} software package, which significantly facilitated our analysis. Further, we acknowledge the MINOT software for obtaining the spectral model. Authors would like to thank Remi Adam and Vardan Baghmanyan for the useful discussions. We also extend our gratitude to the IUCAA for providing the resources and research environment. Finally, we would like to thank the anonymous referee for their valuable comments and suggestions.\\

\section*{Data Availibility}
The data that support the findings of this study are openly available in Fermi-LAT data server {\url{https://fermi.gsfc.nasa.gov/ssc/data/access/}}. The analyzed data can be made available upon reasonable request to the authors.
\appendix

\section{Systematic Effects\label{app11}} We assessed systematic uncertainties on the extended model ($n_{CR_P}\propto n^{1/2}_{gas}$) arising from the Galactic interstellar emission model (IEM), the choice of event class, the low-energy threshold and PSF0-3. We obtained a set of 12 versions of the diffuse Galactic background components \citep{Vladimirov2011}\footnote{See for more details \url{https://galprop.stanford.edu/},and \url{https://galprop.stanford.edu/webrun.php}}(together with the corresponding isotropic diffuse emission modeled simultaneously), which include bremsstrahlung, inverse Compton, and pion-decay emission. The analysis was repeated using both SOURCE and CLEAN event classes, and the energy threshold was varied between 100 and 500 MeV. The resulting systematic uncertainties on the spectral parameters are summarized in Table~\ref{tab:systematics effects}.

\begin{table*}[htbp]
\centering
\caption{{Systematic uncertainties in the Fermi-LAT analysis from variations in analysis assumptions.}}
\label{tab:systematics effects}
\begin{tabular}{lccc}
\hline \\
\textbf{Type} & \textbf{Details} & \textbf{variation in TS } & \textbf{impact on Flux } \\ \\
\hline \\
Diffuse background       & Alternative models       & 15-21   & $< 20\%$ \\ [1ex]
Energy threshold         & 100--500 MeV             & 17-20   & $<30\%$  \\ [1ex]
PSF0-3                    & Combined           & $\sim 20$   & $\approx 36.9$  \\ [1ex]
Event selection          & Alternative selections   & 15-18 & $<15\%$  \\ [1ex]

\hline
\end{tabular}
\end{table*}

\section{MCMC Results\label{app1}}

\begin{figure*}[ht] 
    \subfloat[]{%
        \includegraphics[width=0.5\textwidth]{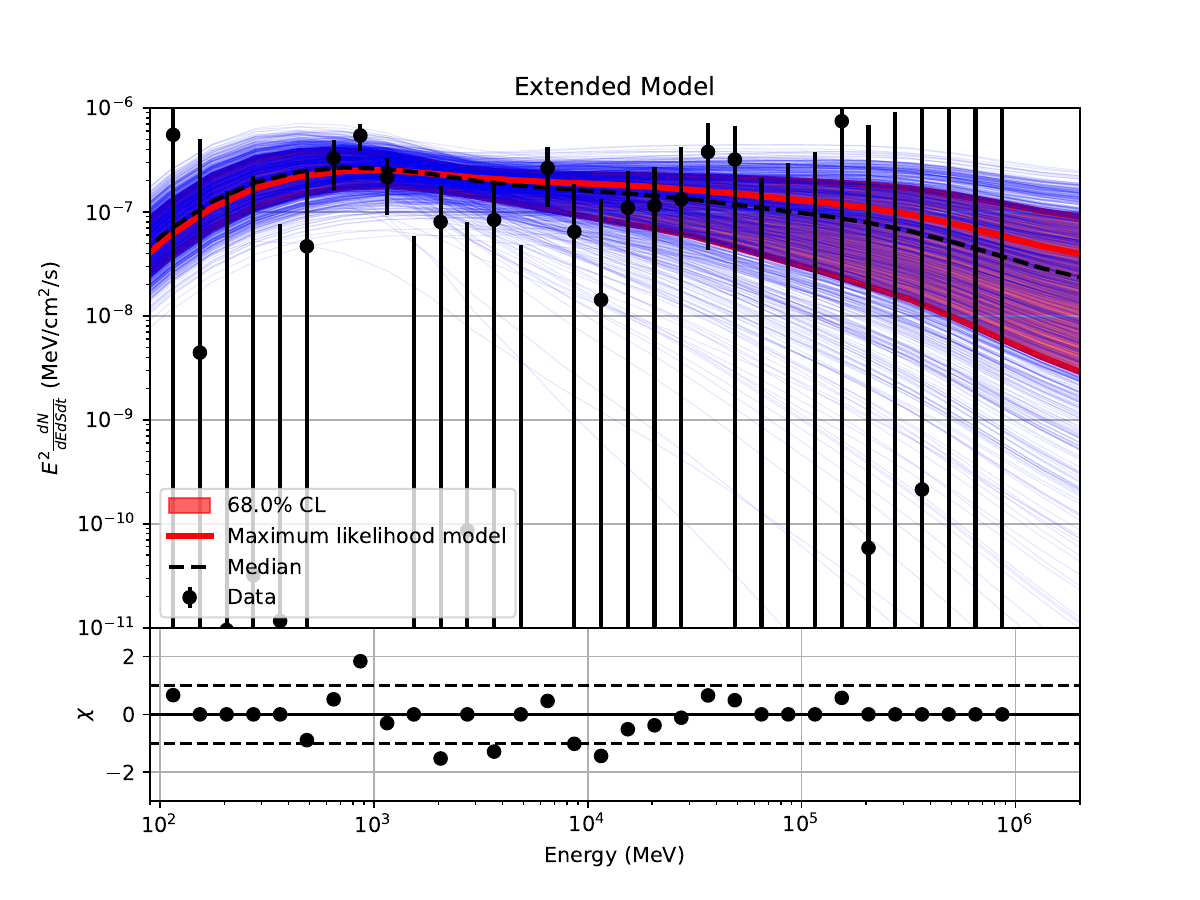}%
        }%
     \hfill%
    \subfloat[]{%
        \includegraphics[width=0.5\textwidth]{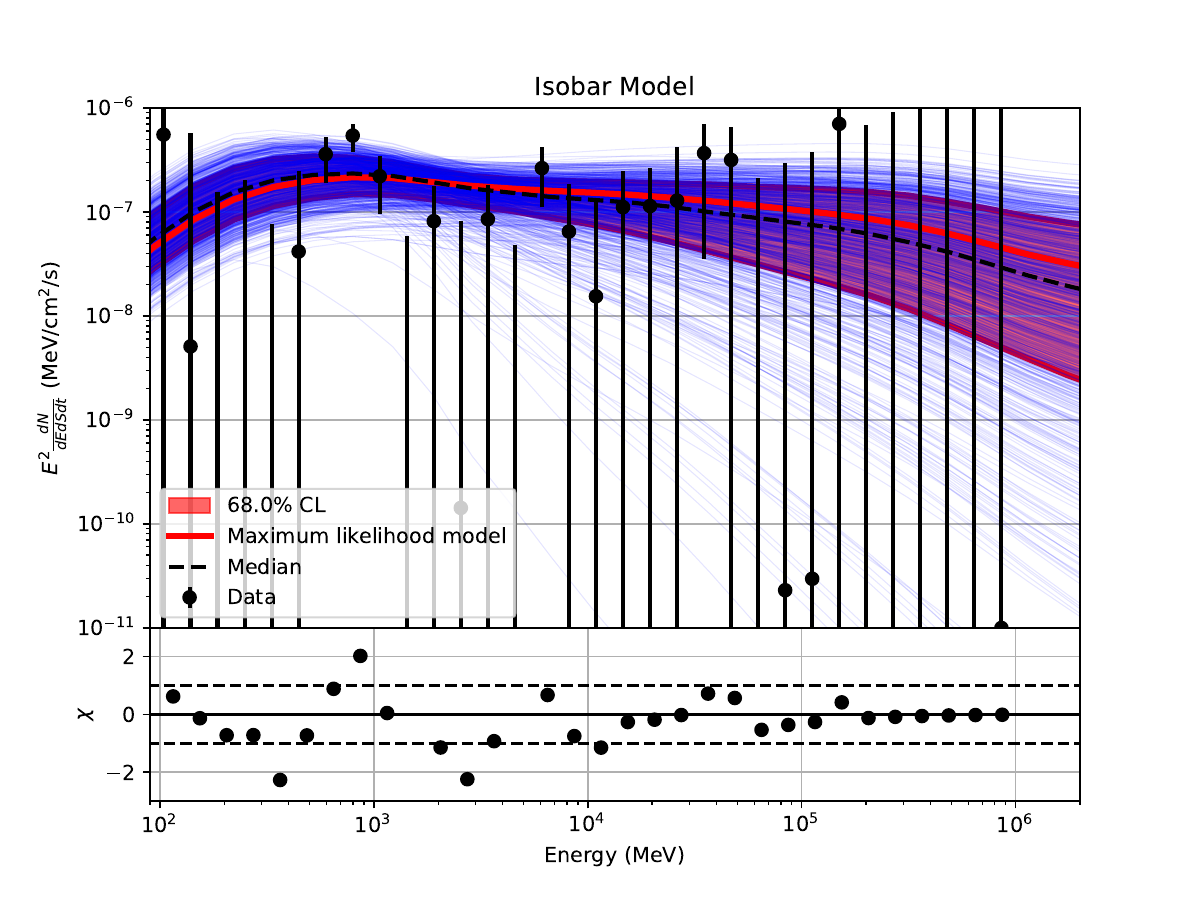}%
        }%
       
    \hfill%
    \subfloat[]{%
        \includegraphics[width=0.5\textwidth]{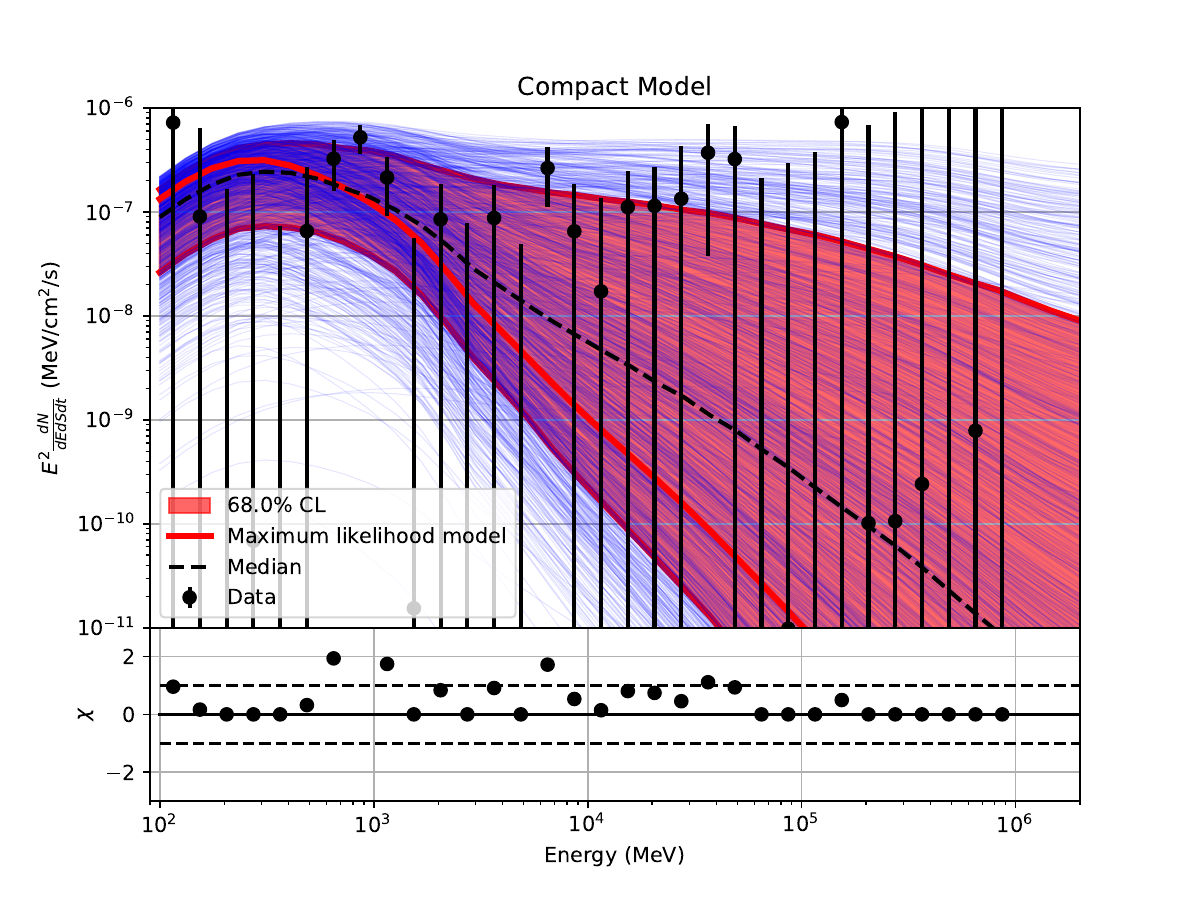}%
        }%
      \hfill%
     \subfloat[]{%
        \includegraphics[width=0.5\textwidth]{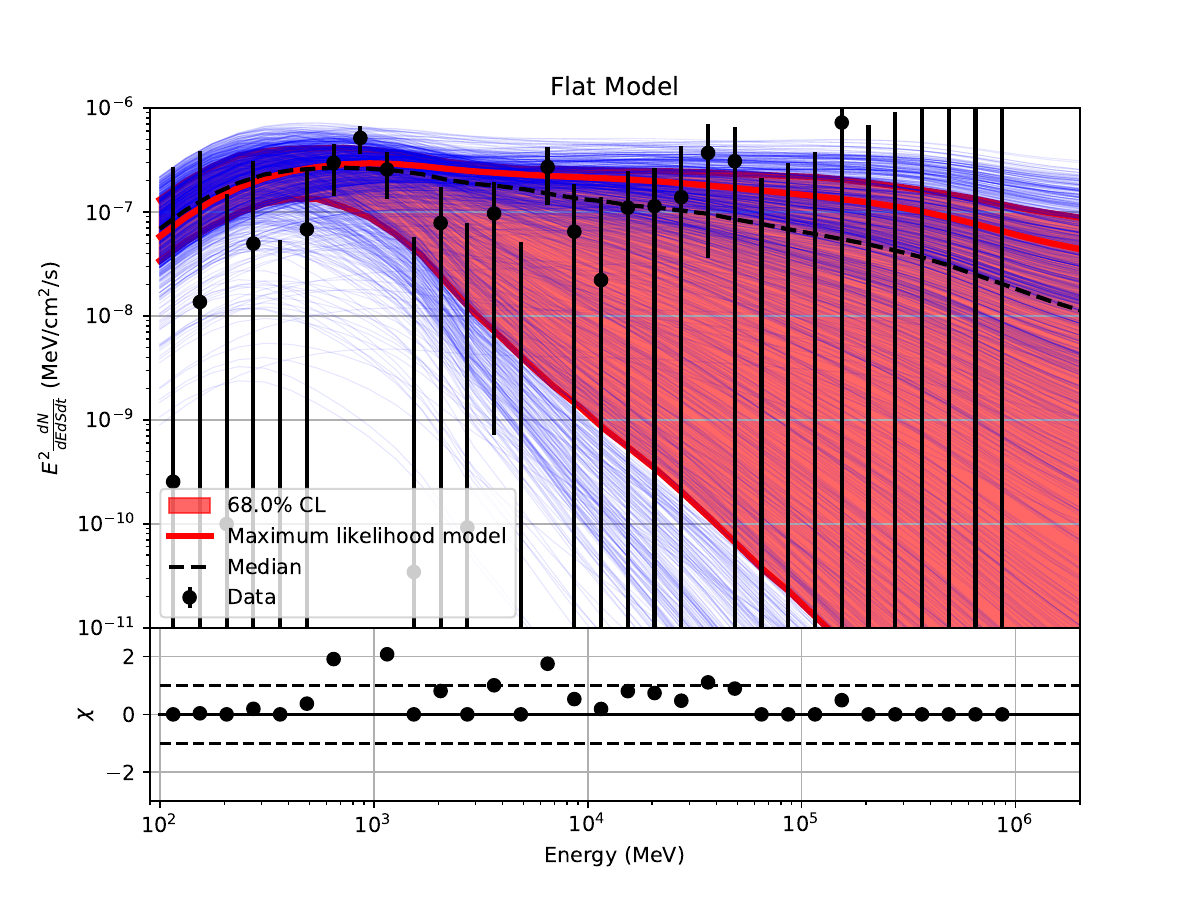}%
         }%
        
    \caption{With free Background: The resultant Spectral Energy Distributions (SEDs) for A119, spanning an energy range from 100 MeV to 1 TeV, obtained with the \textit{sed} function of \textit{Fermipy}. From left to right and top to bottom: (1) Extended Model, (2) Isobar Model, (3) Compact Model, and (4) Flat Model, each fitted using MCMC. Each panel displays the best-fit SED for the respective model. The thick red line indicates the maximum likelihood model, and the light red region indicates the $68\%$ confidence level along with 1000 models obtained using MCMC simulation.The dash black line shows the median of these models. The residuals shown here are the difference between data and the best fit model, normalized with error-bars.}
    
    \label{fig:SED_with mcmc_free_background}   
 \end{figure*}
\begin{figure*}[ht] 
    \subfloat[]{%
        \includegraphics[width=0.45\textwidth]{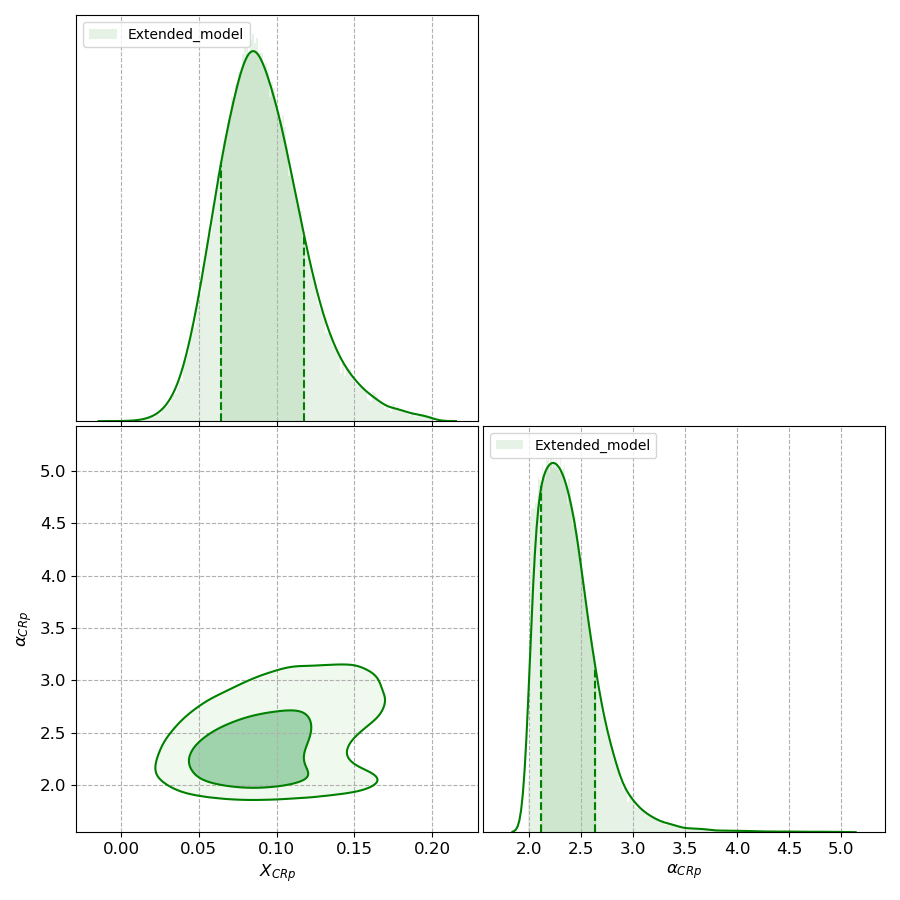}%
        }%
    \subfloat[]{%
        \includegraphics[width=0.45\textwidth]{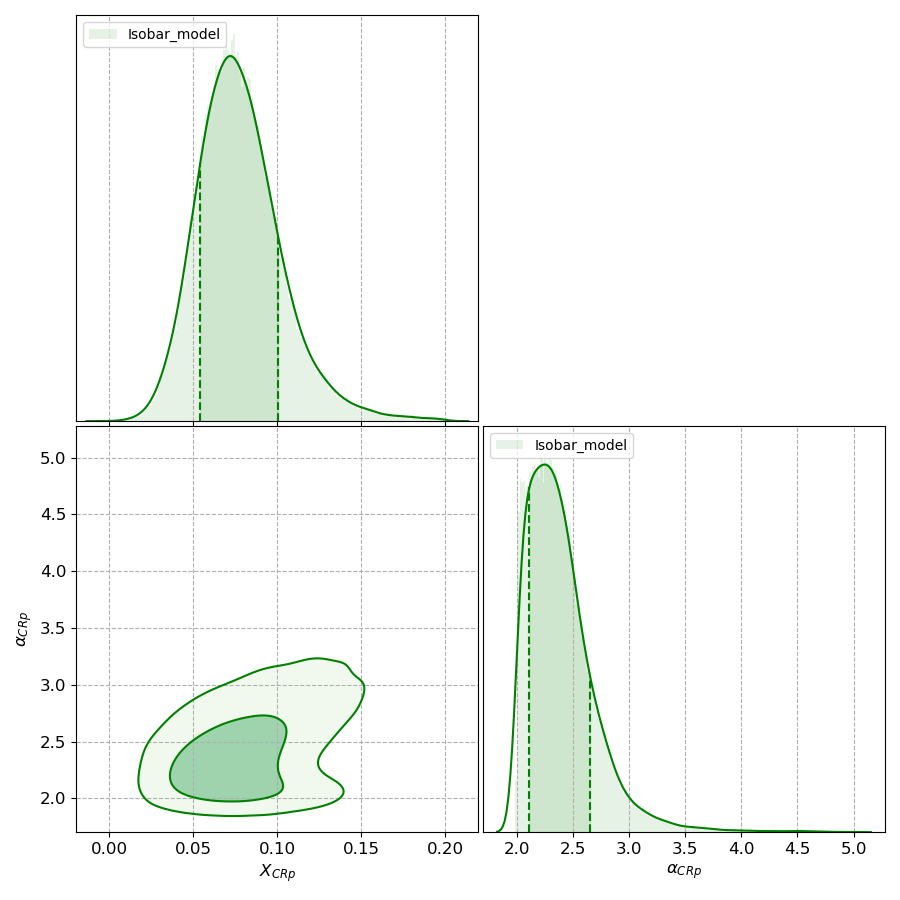}%
        }%
         \hfill%
    \subfloat[]{%
        \includegraphics[width=0.45\textwidth]{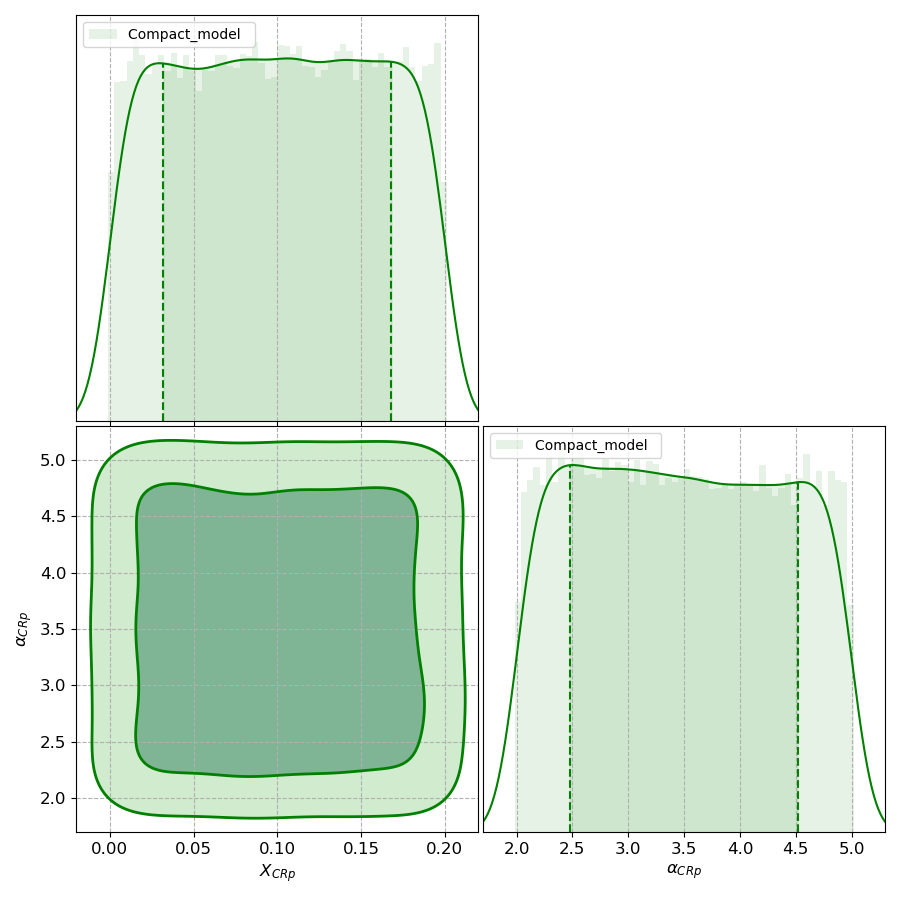}%
        }%
    \centering       
     \subfloat[]{%
        \includegraphics[width=0.45\textwidth]{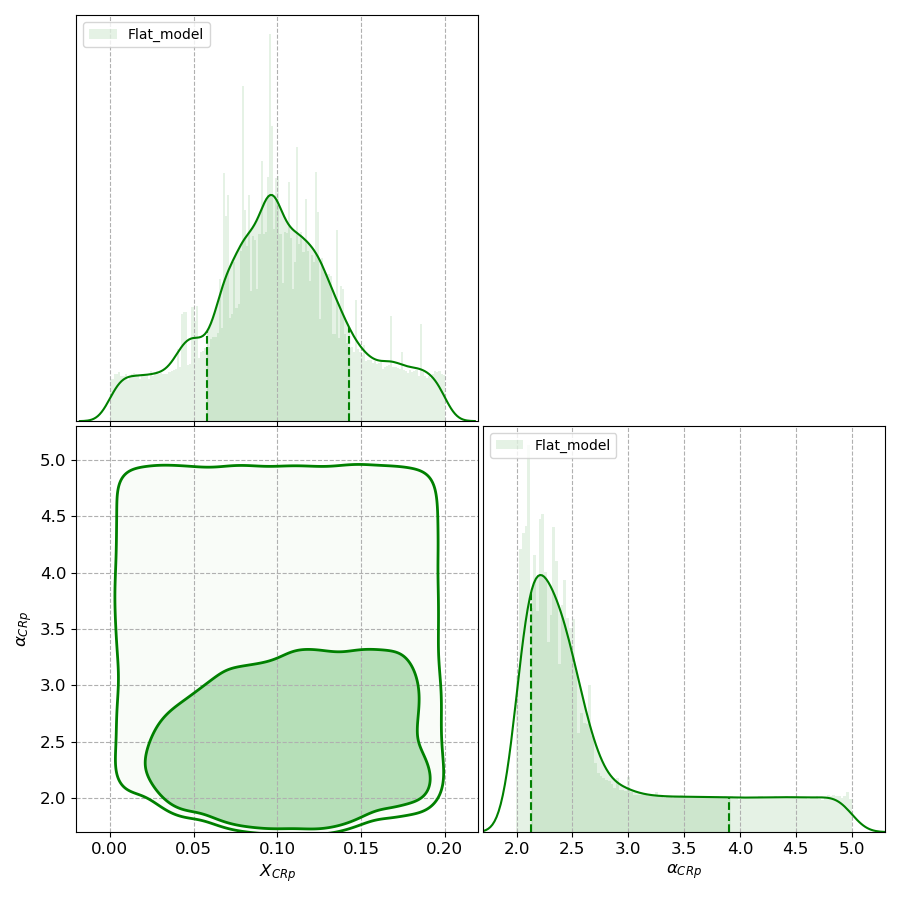}%
         }%
           
        \caption{ Free Background: MCMC models constraints, displayed as corner plots. From left to right and top to bottom: (1) Extended Model, (2) Isobar Model, (3) Compact Model, and (4) Flat Model. Each plot presents the posterior distributions and parameter correlations, reflecting the MCMC sampling results for each model scenario. The posterior probability distribution is shown at the top for the $X_{CRp}(R_{500})$ and at the bottom right for the $\alpha_{CRp}$, where the green shaded area indicates the $68\%$ confidence interval. The bottom left panel shows the contour levels at the $68\%$ confidence interval in the dark green shaded area and the $95\%$ confidence interval in the light green shaded area for the $X_{CRp}(R_{500}) - \alpha_{CRp}$.}
    \label{fig:corner_freebackground}    
 \end{figure*}

 
The posterior likelihood constraints for $X_{CRp}(R_{500})$ and $\alpha_{CRp}$ of each model are shown in \autoref{fig:corner_freebackground} while in the case of free background.

\subsection{free Background :Luminosity Estimation}
\autoref{tab3_freebackground} presents the MCMC constraints on $X_{CRp}(R_{500})$ and $\alpha_{CRp}$ across all tested spatial models. Additionally, the table includes the fluxes and corresponding luminosities, as derived from the MCMC fits for each model, providing a comprehensive view of the CRp population's impact on the A119, diffuse $\gamma$-ray emission.
\begin{table*}[ht!]
\caption{Free Background: Constraints on the CRp Population, Flux, and Luminosity for CRp Models — This table presents the MCMC-derived constraints on the CRp normalization $X_{CRp}(R_{500})$ and spectral slope $\alpha_{CRp}$ along with the associated $\gamma$-ray flux and luminosity values, for each tested CRp model configuration. \label{tab3_freebackground}}
    \begin{tabular}{ccccc}
    \toprule \\
ICM or Sky Model       &  $X_{CRp} $     &  $\alpha_{CRp}$  & Flux ($10^{-10}$ ph cm$^{-2}$ s$^{-1}$ )  & Luminosity ($10^{42}$ erg s$^{-1}$) \\ \\
\midrule
Extended Model($n_{CR_P}\propto n^{1/2}_{gas}$)&  $0.077^{+0.04}_{-0.01}$  &$2.25^{+0.38}_{-0.13}$  &$13.94^{+10.26}_{-4.43}$  & $11.49^{+3.52}_{-3.33}$  \\ \\
Isobar Model($n_{CR_P}=P_{gas}$)  &$0.065^{+0.04}_{-0.01}$ &$2.26^{+0.39}_{-0.15}$  &$11.97^{+9.52}_{-3.90}$  &$9.55^{+2.94}_{-2.64}$   \\ \\
Compact Model($n_{CR_P}\propto n_{gas}$)  & $0.13^{+0.03}_{-0.10}$   & $4.21^{+0.315}_{-1.74}$ & $21.39^{+9.96}_{-1.61}$ &$4.69^{+7.90}_{-3.21}$   \\ \\
Flat Model($n_{CR_P}=$Constant) & $0.09^{+0.05}_{-0.03}$  &$2.28^{+1.64}_{-0.12}$  &$16.18^{+11.00}_{-6.68}$  & $13.17^{+2.86}_{-9.49}$ \\ \\

\bottomrule 

\end{tabular}
\begin{tablenotes}

\item \textbf{Note:} The reported values and uncertainties represent the maximum likelihood estimates along with the $68\%$ confidence intervals derived from the distributions.

\end{tablenotes}
\end{table*}

\bibliographystyle{mnras}
\bibliography{bibtex_prd}

\def\apj{ApJ}%
\def\mnras{MNRAS}%
\def\aap{A\&A}%
\def\apjl{ApJ}
\def\aj{AJ}
\def\physrep{PhR}
\def\apjs{ApJS}
\def\pasa{PASA}
\def\pasj{PASJ}
\def\nat{Natur}
\def\apss{Ap\&SS}
\def\araa{ARA\&A}
\def\aaps{A\&AS}
\def\ssr{Space Sci. Rev.}
\def\pasp{PASP}
\def\na{New A}
\def\jcap{JCAP} 
\def\aap{AAP}

\end{document}